\documentclass{aa}

\usepackage{graphicx}

\usepackage{txfonts}

\usepackage{natbib}
\bibpunct{(}{)}{;}{a}{}{,}

\begin{document}

   \title{Connecting the long-term variability behaviour of active galactic nuclei to their central engines}

   \author{S. Kankkunen \inst{1,2}, M. Tornikoski \inst{1}, T. Hovatta \inst{1,2,3}
          }

   \institute{Aalto University Metsähovi Radio Observatory, Metsähovintie 114, FI-02540 Kylmälä, Finland
              \\
              \email{sofia.kankkunen@aalto.fi}
                \and
            Aalto University Department of Electronics and Nanoengineering, PO Box 15500, 00076 Aalto, Finland
                \and
            Finnish Centre for Astronomy with ESO, FINCA, University of Turku, Turku, FI-20014 Finland
            }
    \date{Received November 11, 2025; accepted March 17, 2026}

  \abstract
   {}
   {Analysing the long-term radio variability of active galactic nuclei (AGNs) is essential to understanding the physics of relativistic jets launched by supermassive black holes. We aim to connect the characteristic timescales obtained from a prior power spectral density (PSD) analysis to the decomposed timescales of the light curves. In addition, we probe for potential associations between the timescales and the physical characteristics of the relativistic jet as well as the central engine.}
   {We decomposed the long-term radio light curves of 54 sources observed at the Aalto University Metsähovi Radio Observatory into individual flares to understand which timescale of variability is related to the low-frequency bend in the PSD. In addition, we used the obtained rise times of the brightest flares to look for associations between the emission-region size in the jet and different central engine parameters.
   }
   {We found that the inverse of the PSD bend frequency of radio light curves best corresponds to the mean duration of the brightest flares. For some sources, the mean flare separation had a similar timescale. Using the flare durations and separations as proxies for the PSD timescale, we found a positive correlation with black hole mass divided by the normalised mass accretion rate. This suggests that the variability timescales obtained from the PSDs of radio light curves are associated with the central engine. Furthermore, when comparing the obtained rise times of the brightest flares to the jet and central engine parameters, we found weak tentative correlations, but they may be driven by a common dependency on redshift.} 
   {}

   \keywords{galaxies: active --
                quasars: general --
                methods: data analysis
               }
\titlerunning{Connecting the long-term variability behaviour of AGN to their central engines}
   \maketitle

\section{Introduction}

Active galactic nuclei (AGNs) are highly luminous centres of active galaxies and are known for their multi-wavelength variability. In the radio domain, the variability has been attributed to shocks in jets \citep{marscher1985models}, where the observed luminosity and rate of variability are boosted by relativistic flow velocities and the jet orientation being close to the line of sight. The continuum observations of flares have been connected to very long baseline interferometry (VLBI) observations of ejected features in the jet (e.g. \citealt{turler1999modelling}; \citealt{savolainen2002connections}; \citealt{lindfors2006synchrotron}). 

The time variability of AGNs has been studied extensively in all wavelength domains (e.g. \citealt{hughes1992university}; \citealt{uttley2002measuring}; \citealt{mchardy2006active}; \citealt{hovatta2007statistical}; \citealt{kelly2009variations}; \citealt{ramakrishnan2015locating}). Typically, variability studies attempt to look for so-called characteristic timescales of variability. A characteristic timescale describes the typical variability in the light curves, such as how often the source flares and for how long the flaring lasts. These timescales may be connected to the physical conditions in AGNs, and they are an important piece in understanding the drivers of AGN variability over different wavelengths. 

One typical way to estimate characteristic variability from observed light curves is to analyse their power spectral densities (PSDs). The PSD gives the power content of a signal (or noise) in frequency space, and it can be estimated using the periodogram. The PSD of an AGN is typically modelled as a red noise power law with a potential bend after which the slope turns over to a flatter power-law slope or converges to white noise in low frequencies. This observed low-frequency flattening is expected in real physical processes to avoid infinite integrated power (e.g. \citealt{van1989quasi}).

A bend in the bending power law is considered to be the inverse of a physically meaningful characteristic timescale; in black hole X-ray binaries (BHXRBs), the emergence of jets is associated with changing X-ray spectral states (e.g. \citealt{fender2004towards}), and their light-curve PSDs have either one or two observable bends (e.g. \citealt{belloni1990variability}; \citealt{mchardy2006active}; \citealt{wilkinson2009accretion}). The location of the bend has been connected to central engine parameters of both BHXRBs and AGNs (e.g. \citealt{uttley2002measuring}; \citealt{papadakis2004scaling}; \citealt{mchardy2006active}; \citealt{gonzalez2012x}; \citealt{paolillo2023universal}). Specifically, \cite{mchardy2006active} found the X-ray PSDs of radio-quiet AGNs to be similar to the soft-state PSDs of BHXRBs with a single bend. They found the location of the PSD bend to scale with the black hole mass and mass accretion rate. This scaling was also found to apply to the PSD of the blazar 3C273 \citep{mchardy2009x}. Thus, of interest is to analyse whether the PSD bend of radio light curves can also be associated with the central engine.

In \cite{kankkunen2025long} (hereafter Paper I), we were able to identify 11 sources for which there was a constrainable characteristic timescale. Due to the small sample of sources with a PSD bend, we attempt to extract the characteristic timescale from the light curves via light-curve decomposition. We test the hypothesis that the timescale we obtain from the PSD, the inverse of the bend frequency, is connected to either the durations of flares or to their temporal separations. The idea that the longest timescales are related to the flare durations is supported by the preliminary analysis in Paper I, where we compared the PSD timescales to the longest times a knot component was visible in very long baseline interferometry (VLBI) 43 GHz data \citep{weaver2022kinematics}. Similarly, \cite{park2017long} found that introducing longer-duration flares to their simulated light curves moved the bend location to lower frequencies. On the other hand, \cite{mukherjee2019accretion} used simulated data and light curves to analyse the connection between the PSD timescale and the light-curve characteristics and found the PSD bend to be the inverse of the intervals between consecutive shocks moving down the jet. Thus, it is not clear, which timescale observed in the light curve corresponds best to the PSD bend, and different timescales may also be blended. Nevertheless, if a connection is found, we can use the corresponding timescale as a proxy for the PSD bend and test for a connection with the central engine with a larger sample of sources.

In addition to the connections between the PSD timescale and the central engine, we analyse whether the jet and central engine parameters can be related to the emission region size. For example, a study by \cite{potter2015new} found compelling evidence of a connection between the bulk Lorentz factor of the jet and the radius of the jet transition region. The transition region is defined as the region where the jet shape changes from parabolic to conical and is in equipartition. A transition between the jet shapes has been observed in multiple AGNs (e.g. \citealt{asada2012structure}; \citealt{kovalev2020transition}; \citealt{boccardi2021jet}; \citealt{okino2022collimation}). \cite{kovalev2020transition} suggest that the transition region likely coincides with the location of a recollimation shock. In millimetre wavebands, the radio core is often considered to be a standing recollimation shock (\citealt{daly1988gasdynamics}), although in some sources it may still be the \(\tau\) = 1 surface (e.g. \citealt{kutkin2014core}; \citealt{fromm2015location}). With simplifications and assuming a direct relationship between the length of the parabolic region and the width of the jet shape transition region (e.g. \citealt{vlahakis2003relativistic}; \citealt{beskin2006effective}; \citealt{lyubarsky2009asymptotic}; \citealt{potter2015new}), we incorporate light-travel time arguments to use the decomposed flares and their emission-frame rise times as proxies for the width of the radio core. We estimate whether the rise times can be related to the jet bulk Lorentz factor, the black hole mass, the accretion disk luminosity, and/or the mass accretion rate.
 
Our sample includes both the constrained sources of Paper I as well as a larger sample of additional sources for which a constrained timescale was not found. Comparing the results between the decompositions and the PSD timescales, or lack thereof, should help in further understanding the characteristic variability and its dependence on jet physics. The paper is divided as follows: In Section 2, we describe the used data and sample. In Section 3, we describe the methods, followed by the results in Section 4. In Section 5 we discuss the results and their connection to previous studies. Section 6 gives the conclusions obtained from the analysis. 

\section{Data}

The Aalto University Metsähovi Radio Observatory (MRO) hosts a 14-metre telescope that is used to observe AGNs predominantly in 37 GHz. Observations have been conducted regularly for over 40 years, with a few sources having their first observation already in 1979. Because of this long-term monitoring, the MRO sample of light curves allows us to analyse the time variability of the most energetic relativistic jets with high accuracy. More detailed information on MRO and the observed sample is presented in \cite{terasranta1998fifteen} and in Paper I.

In Paper I, we compiled a sample of 123 sources from the objects monitored at MRO, which had a minimum of 10 years of consecutive observations, a minimum of 100 data points, and at least a 1 Jy maximum flux density. For this work, we constructed a subsample by including the 11 constrained sources with a constrained timescale obtained from the PSD analysis, and a subset of 40 well-sampled unconstrained sources as a further comparison sample. We also included three borderline sources from Paper I, which are sources that had a timescale in the PSD but did not meet all strict criteria to be considered reliable. The additional sources in the analysis were chosen from Paper I with the following criteria: a minimum of 500 data points, no multiple long gaps in the light curve, and prominent flares for the decomposition analysis. Including this larger sample of sources should allow for further confirmation of their timescales and potentially relate the results to the central engine of the sources. The entire sample consists of 54 sources and they are listed in Table \ref{table:sources}. 

Because we are interested in comparing characteristic timescales obtained from the PSD to the light-curve decomposition, we used the same cut-off date as in Paper I (1 January 2023), leading to a median light-curve length of \(\sim\)39 years. The source types in the sample can be divided based on optical classification into flat-spectrum radio quasars (FSRQs), BL Lac objects (BLOs), and radio galaxies (GALs).

\section{Methods}

\subsection{Light-curve decomposition} \label{lcdecomp}

To connect our results from the PSD analysis in Paper I to the specific timescales in the light curves, we applied a decomposition algorithm. Decomposing the light curves into individual flares has been shown to be a viable method for estimating radio light curve parameters (e.g. \citealt{lahteenmaki1999total}; \citealt{hovatta2009doppler}), and it allows us to compare the light curve characteristics to the PSD timescales. Using both the results from the flare decomposition and the PSD analysis may help in understanding whether the light curve is sufficiently long to describe the source behaviour. If no bend is visible in the PSD, it suggests that the amplitude of the long-term variations in the observed light curve is still increasing (e.g. \citealt{press1978flicker}), and thus the behaviour of the source cannot be fully characterised. 

The Magnetron\footnote{https://github.com/dhuppenkothen/magnetron2/tree/blazars/} code \citep{huppenkothen2015dissecting} decomposes light curves into separate flares, with an exponential rise and decay, with no a priori information on the number of flares or the parameters of the exponential function. The algorithm uses Bayesian hierarchical modelling and diffusive nested sampling (DNest4, \citealt{brewer2011diffusive}; \citealt{brewer2018dnest4}) to determine the number of components and the parameters of individual components independently of each other and to create samples from the posterior. For the list of priors and other assumptions used, we refer the reader to \cite{huppenkothen2015dissecting}.
 
As described in detail in \cite{liodakis2018constraining}, the modification of the Magnetron software to blazar analysis uses the Ornstein-Uhlenbeck (OU) process to model the underlying background variability unrelated to flaring events. The OU process is a mean-reverting stochastic stationary process \citep{uhlenbeck1930theory}.  If no OU process is modelled for the underlying large-scale variability, the program tries to fit each small variation in the light curve with an exponential flare. The OU background should thus help in avoiding fitting exponential flares to minor variations due to turbulent processes or to the extended diffuse emission from the jet. 

Figure \ref{fig:Decomposition} shows an example of a single realisation of the decomposition drawn from the posterior sample of source 0415+379. The posterior sample includes several possible decompositions with different numbers of individual flares and varying flare parameters. This way the method accounts also for the uncertainty in the number of flares in any given light curve as often multiple variations of flare superpositions fit the light curve equally well. The number of obtained posteriors varied between $\sim$250-1250, with a median of 517 samples across all sources. For each observed light curve, the plots of the posterior weights were monitored to decide when a run was successfully finished (see \citealt{brewer2018dnest4}). All of the decomposed flares are described by their peak position and amplitude, e-folding timescale (time constant) and skewness. The e-folding timescale is related to the rise of the flare, whereas the decay time is obtained from multiplying the timescale with the skewness factor. 

Upon visual examination of some of the decompositions, we noticed that some posterior draws included suspiciously large spikes and occasionally extremely low flare amplitudes. The Diffusive Nested Sampling algorithm \citep{brewer2018dnest4} used by the Magnetron code, is a Markov chain Monte Carlo method with inherent statistical randomness, so it may be possible that a different run of the code results in a different fit. Therefore, we ran the decomposition on some of the sources twice and found that the results of the test runs were within the standard deviations of the original run. Thus, we concluded that the posterior distributions were sufficiently consistent between different runs. To mitigate the effect of the fast spikes, we required the e-folding timescale of each flare to be longer than the median cadence in the light curve for the flare to be included in the analysis.

\subsubsection{Flare durations and separations}

We estimate whether the PSD timescale can be associated with something easily measurable in the decomposed AGN light curves. It is possible that multiple timescales can be extracted from the PSD; however, in Paper I, we only probed for a single low-frequency bend in the PSD. To find the potential timescale associated with the PSD timescale, we test two simple parameters obtained from the light curves: the flare durations and the separations of their peaks. The separations obtained from the peaks roughly correspond to the intervals between flare onsets, with the exception of a few sources that only include a small number of exceptionally long flares whose onsets are not included within the observing windows.

\begin{figure*}
    \centering
    \includegraphics[width=1\linewidth]{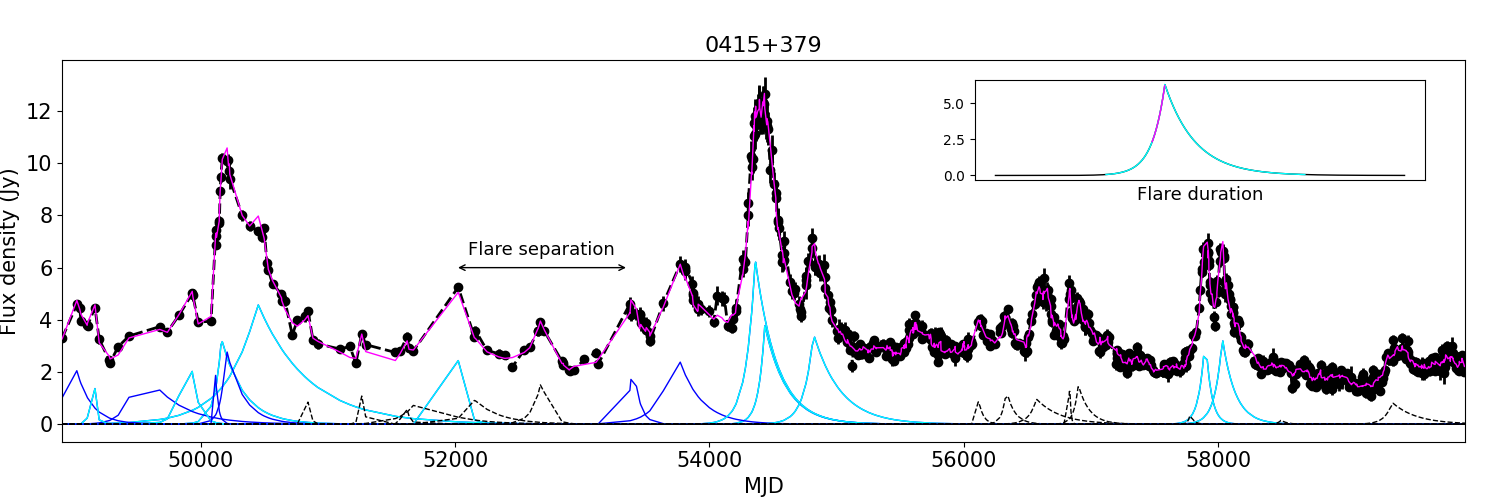}
    \caption{Example of one decomposition of the 37 GHz light curve of 0415+379 drawn from the posterior sample. The OU process has been subtracted from the decomposition. The flares coloured in blue and cyan have an amplitude higher than the mean amplitudes of all flares in the posterior, and the flares coloured in cyan have amplitudes above 50 \% of the maximum amplitude in the given posterior. The magenta line shows the fit of the decomposition. In the top right, the inset shows an example of the portion of the flare used to determine the flare duration (cyan) and the portion of the flare used for the rise time (magenta).}
    \label{fig:Decomposition}
\end{figure*}

Because the red noise nature of AGN PSDs indicates that the amplitude of variations is increasing over longer and longer timescales (e.g. \citealt{press1978flicker}), we focus the analysis on the brightest flares. We analyse those flares that have an amplitude above the mean amplitude in a given posterior draw. This decision allows a sufficient number of flares to be included in analysing both the mean durations and the flare peak separations; however, this criterion may underestimate the maximum separations in case there are only a small number of very bright flares. An example of flare decomposition with the mean-amplitude criterion is given in Fig. \ref{fig:Decomposition}. The figure also shows an example of the flare separation, visualised between two adjacent flares. The durations are determined from when the flare has reached 1 \% of its maximum amplitude on the rise and decay sides. This limit was also used in \cite{huppenkothen2015dissecting} with count rates. To determine the timescales from the durations and separations of the most dominant flares, we took their means for each posterior, and then similarly the means from all drawn posteriors. 

\subsubsection{Flare rise times}

The brightest flares have been linked to so-called core flares associated with ejections of VLBI components \citep{savolainen2002connections}. As a simplified assumption, the rise times of the brightest flares should give the best representation of variability timescales within the radio core. We exclude the fainter flares as they have been associated with knots further downstream of the core (e.g. \citealt{savolainen2002connections}). With the following reasoning and light-travel time arguments, we use the e-folding timescale of the most dominant flares as a proxy for the upper limit of the relative emission region size. The e-folding timescale corresponds to approximately 37 \% of the flare maximum amplitude, and we quote it as the rise time. This portion of the flare is indicated in Fig. \ref{fig:Decomposition}.

Here, we narrow down the number of flares included to only consider the ones whose peak amplitudes exceed 50 \% of the highest-amplitude flare in a given decomposition. This is to ensure that a minimal number of flares being formed further downstream are included in estimating the size of the radio core. From each posterior draw, we take the median rise time of the considered flares and then take the median over the sampled posteriors. To scale the rise times to the emission frame, we used

\begin{equation}\label{z_d_correction}
\tau_{\rm rise,em} = \frac{\tau_{\rm rise,obs} \cdot \delta_{\rm var}}{1+z},
\end{equation}
where \(\tau_{\rm rise,obs}\) is the decomposed rise time in the observed frame, \(\delta_{\rm var}\) the variability Doppler factor, and $z$ the redshift.

To be able to compare the rise times of the flares and thus the sizes of the emitting regions between sources, we made several simplifications in our assumptions. First, we assume that synchrotron opacity has a negligible effect on the flare timescales. In addition, to compare the obtained results to previous studies, we assume that the emitting region corresponds to the location of the jet transition region and that the upper limit of the emitting region size is a sufficient proxy for the jet width at that location.

\subsection{Doppler factors}

\cite{lahteenmaki1999total} discussed different methods for estimating Doppler factors. One approach involves directly measuring the brightness temperatures from VLBI data and comparing them to the theoretical maximum equipartition brightness temperature \citep{readhead1994equipartition}. Another option is to use total flux density flares through light-curve decomposition (e.g. \citealt{terasranta1994brightness}; \citealt{lahteenmaki1999total}; \citealt{hovatta2009doppler}), where the intrinsic brightness temperature \(\mathrm{T_{b,int}}\) is compared to the variability brightness temperature \(\mathrm{T_{b,var}}\).

We estimate the Doppler factors following the prescription in \cite{liodakis2021identifying}, using the cosmological parameters \(\Omega_m = 0.315, \Omega_{\Lambda} = 0.685\), and \(\mathrm{H_o=67.4 km s^{-1} Mpc^{-1}}\) (\citealt{aghanim2020planck}) as

\begin{equation}\label{Tbvareq}
T_{\rm b, var} = 1.47 \cdot 10^{13} \dfrac{d_{\rm L}^2\Delta S_{\rm ob}( \nu)}{\nu^2t_{\rm var}^2(1+z)^4}K,
\end{equation}
where \(d_{L}\) is the luminosity distance in Mpc, \(\Delta S_{\rm ob}(v)\) the flare amplitude in Jy, \(\nu\) the observing frequency in GHz, \(t_{\rm var}\) the flare rise time in days, and z the redshift (\citealt{liodakis2021identifying}). From each posterior draw, the maximum \(\mathrm{T_{\rm b, var}}\) is chosen. We use the e-folding timescale as the rise time and consequently the difference between the maximum flux density of a flare and the flux density at the e-folding timescale as \(\Delta S_{\rm ob}(v)\). To account for the uncertainty in \(\mathrm{T_{\rm b,int}}\), we follow \cite{liodakis2018constraining} and use a Gaussian distribution with mean \(\mu\) = 2.78 \(\cdot 10^{11}\)K and standard deviation \(\sigma\) = 0.72\(\cdot10^{11}\) K and draw samples from it. The variability Doppler factor \(\delta_{\rm var}\) was obtained from

\begin{equation}\label{Dopplereq}
\delta_{\rm var} = (1+z)\sqrt[3]{\frac{T_{\rm b, var}}{T_{\rm b, int}}}
\end{equation}
by random sampling from the distribution of \({T_{\rm b, var}}\) and \({T_{\rm b, int}}\) and constructing the median. The \(\delta_{\rm var}\) and the respective 68 \% confidence intervals are reported in Table \ref{table:sources}. Two sources, 3C84 and ON231, had a Doppler factor less than one, suggesting that their intrinsic brightness temperature is likely lower than indicated in the distribution derived in \cite{liodakis2018constraining}. Thus, we excluded the values from the table. We also note that the Doppler-factor estimate for 0836+710 is high, with a value \(\delta_{\rm var}\) = 78.6, and it is possibly an artefact of the decomposition procedure due to some gaps in the beginning of the light curve.

\subsection{Lorentz factor}

We derived the Lorentz factors for each source through the calculated Doppler factors and the apparent jet speeds (\(\beta_{app}\)) from the MOJAVE survey (\citealt{lister2018mojave}; \citealt{lister2019mojave}) using

\begin{equation}\label{Lorentzeq}
\Gamma = \dfrac{\beta_{\rm app}^2+\delta_{\rm var}^2+1}{2\delta_{\rm var}}.
\end{equation}
As there is no reported redshift for 0716+714 in \cite{lister2019mojave}, we adopt z = 0.2304 from \cite{pichel2023statistical}. For S20109+22, we used z = 0.49 \citep{koljonen2024galaxy}. 

Through Eq. \ref{Dopplereq} and Eq. \ref{Lorentzeq}, we were able to estimate the Lorentz factor for 50 sources, reported in Table \ref{table:sources}. The given apparent speeds in units of c have been converted using \(\Omega_m = 0.315, \Omega_{\Lambda} = 0.685\), and \(\mathrm{H_o=67.4 km s^{-1} Mpc^{-1}}\) (\citealt{aghanim2020planck}).

\subsection{Central engine parameters}

To compare the size of the emission region to the central engine parameters of the sources, we estimated the accretion disk luminosities, \({L_{\rm acc}}\), and black hole masses, \({M_{\rm BH}}\), of the sources mainly using broad-line luminosities (\citealt{zamaninasab2014dynamically} and references therein). When available, we use the \(\mathrm{H\beta}\) line for the accretion disk luminosity, otherwise the \(\mathrm{MgII}\) line or the \(\mathrm{OIII}\) line. 

\subsubsection{Black hole mass}

The observed optical continuum emission of radio-loud AGNs includes synchrotron emission from the jet; thus, we needed to use the broad-line luminosities to obtain an estimate for the disk emission. This was done through a power-law relationship between the line luminosities and the continuum, which is constructed through observations of radio-quiet sources. To obtain an estimate for the \(\lambda L_{\rm 5000Å}\) continuum emission, we used the formula from \cite{liu2006jet} and for the \(\lambda L_{\rm 3000Å}\) continuum emission from \cite{shen2011catalog}. The estimate for \(\lambda L_{\rm 3000Å}\) was measured from total line luminosities, but according to \cite{shen2011catalog}, the power-law relationship is similar enough that broad emission lines can be used. We only substituted the observed continuum emission with the estimate from the broad emission lines, if the line estimate gives a smaller luminosity. 

The formulae for an estimate of \({M_{\rm BH}}\) are \citep[][and references therein]{zamaninasab2014dynamically}

\begin{equation}\label{HbetaFWHM}
M_{\rm BH} \simeq 4.82\left[\dfrac{\lambda L_{\rm 5100Å}}{10^{44}\rm erg s^{-1}}\right]^{0.69}\left[\dfrac{\rm FWHM_{H\beta}}{\rm km s^{-1}}\right]^2M_\odot,
\end{equation}

\begin{equation}\label{MgFWHM}
M_{\rm BH} \simeq 3.37\left[\dfrac{\lambda L_{\rm 3000Å}}{10^{44}\rm erg s^{-1}}\right]^{0.47}\left[\dfrac{\rm FWHM_{MgII}}{\rm km s^{-1}}\right]^2M_\odot,
\end{equation}
where \(\mathrm{FWHM_{H\beta}}\) is the width of the \(\mathrm{H\beta}\), and \(\mathrm{FWHM_{MgII}}\) is the width of the MgII emission line.

The majority of the emission lines and full width at half maximums (FWHMs) are obtained from \cite{torrealba2012optical}, supplemented by values from \cite{shaw2012spectroscopy}, \cite{stickel1993complete}, and \cite{cohen1987nature}. We obtained an estimate for the black hole mass for 30 sources; the references for the emission lines are reported in Table \ref{table:sources}. We do not include black hole masses derived through other methods to avoid any additional biases in the analysis.

\subsubsection{Accretion disk luminosity and mass accretion rate}

The formulae to calculate the accretion disk luminosity \({L_{\rm acc}}\) from emission-line luminosities are \citep[][and references therein]{zamaninasab2014dynamically}

\begin{equation}\label{Hbeta}
\log_{10}L_{\rm acc} = (12.32 \pm0.32) + (0.78\pm0.01)\rm log_{10}L_{\rm H\beta},
\end{equation}

\begin{equation}\label{MgII}
\log_{10}L_{\rm acc} = (16.76 \pm0.26) + (0.68\pm0.01)\rm log_{10}L_{\rm MgII},
\end{equation}

\begin{equation}\label{OIII}
\log_{10}L_{\rm acc} = (26.50 \pm0.32) + (0.46\pm0.01)\rm log_{10}L_{\rm OIII}.
\end{equation}

The mass accretion rates can be calculated from the disk luminosities following the relation for thin disks \citep{shakura1973black}:

\begin{equation}\label{Lacc}
\dot M = \dfrac{L_{\rm acc}}{\eta c^2},
\end{equation}
where \(\dot M\) is the mass accretion rate, \({L_{\rm acc}}\) the accretion disk luminosity, and \(\eta\) the radiative efficiency. We adopt the value \(\eta\) = 0.4, similarly to \cite{zamaninasab2014dynamically}, albeit it is somewhat unimportant in the upcoming analysis as the efficiency is kept the same for each source. It is noteworthy that this formula is intended for thin disks, and as such it may not work for BLOs. Nevertheless, because we are using the same scaling factor for each source, the mass accretion rate will have the same relationship with the observed values as the accretion disk luminosity. We obtained the accretion disk luminosity for 34 sources. 

We calculated the normalised mass accretion rates using the accretion disk luminosity and the black hole mass through the formula 

\begin{equation}\label{dotm}
\dot m = \dfrac{\dot M}{\dot M_{\rm Edd}}=\dfrac{L_{\rm acc}}{\eta L_{\rm Edd}},
\end{equation}
where \(L_{\rm Edd}\) \(\approx 1.3 \cdot 10^{47} \frac{M_{\rm BH}}{10^9 M_\odot}\) is the Eddington luminosity (e.g. \citealt{bottcher2012relativistic}). 

We note that the accretion disk luminosities obtained through the methods discussed in \cite{zamaninasab2014dynamically} give a larger value than those obtained through spectral energy distribution (SED) modelling (\citealt{ghisellini2011transition}; \citealt{ghisellini2014power}). This also results in the normalised mass accretion rates being relatively large. However, this should not affect the results of our correlation analysis as the values are calculated in the same way for each source.

\section{Results}

We decomposed the light curves of 54 sources to estimate the mean flare duration and separation of the brightest flares of each source. These timescales are given in Table \ref{table:timescales}.

We used the Kendall's \(\tau\) rank correlation to test for monotonic associations between parameters. We opted for a typical significance level of 0.05, but we also report the exact p-values. We considered a correlation weak if 0.1 < \(\tau\) < 0.3, moderate if 0.3  \(\leq\tau\) < 0.6, and strong if \(\tau \geq\)  0.6. 

\subsection{Flare durations and separations}

\begin{figure*}
    \centering
    \includegraphics[width=1\linewidth]{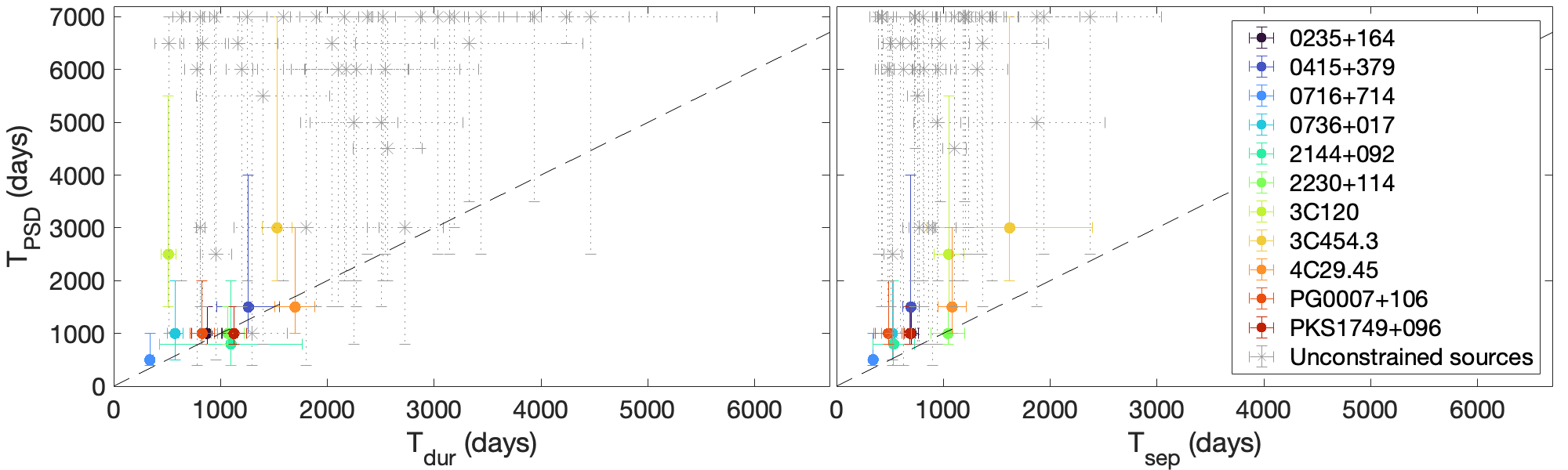}
    \caption{Power spectral density (PSD) timescales against the mean flare durations and separations. The coloured data points are for the constrained sources and the grey ones for the unconstrained sources. The dashed line indicates the one-to-one correspondence between the timescales. }
    \label{fig:PSDDurSep}
\end{figure*}

\begin{table} 
\renewcommand{\arraystretch}{1.3}
\caption{Results from the Kendall rank correlation analysis for associations between the PSD timescale and flare durations and separations.}         
\centering          
\begin{tabular}{l c c c l l l }    
\hline\hline                     
$T_{\rm PSD}$ & \(\tau\) & \(p\) &\\ 
\hline                    
   $T_{\rm dur, all}$ & 0.379 & \(2.803\cdot10^{-4}\) \\
   $T_{\rm sep, all}$& 0.279 & 0.008 \\
   $T_{\rm dur, const}$ & 0.366 & 0.161 \\
   $T_{\rm sep, const}$ & 0.691 & 0.007 \\
\hline           
\end{tabular}
\label{table:psdvsdursep}
\tablefoot{\(T_{\rm dur, all}\) (mean flare duration) and \(T_{\rm sep, all}\) (mean flare separation) include all sources with an estimated PSD timescale and \(T_{\rm dur, const}\) and \(T_{\rm sep, const}\) include only the constrained sources.}
\end{table}

In Paper I, the characteristic timescales were fitted with a bend-timescale grid of 500 days up to 7000 days, with an additional coverage of 100, 200, 400, and 800 days. The coverage was chosen to be sparse because it is difficult to extract exact values from the periodogram, and the uncertainties were correspondingly large. Figure \ref{fig:PSDDurSep} shows the results from the flare decomposition against the obtained PSD timescales. The plots suggest that the flare durations have a better overall one-to-one agreement with the PSD timescale. To quantify this, we performed the Kendall rank correlation test between the PSD timescales and the flare durations and separations. Because most of the PSD timescales of the unconstrained sources are near or at the upper limit of the tested scale (7000 days), we conducted two tests, one with only the constrained sources and one with all of the sources with an estimated PSD timescale. This analysis specifically probes whether the flare durations and/or separations increase with the PSD timescale. We assumed that for the unconstrained sources the PSD timescale is longer than for the constrained sources. The correlation analysis results are reported in Table \ref{table:psdvsdursep}.

We obtained significant correlations with both the flare durations and separations when using all of the sources. For the 11 constrained sources, the correlation is not significant with the flare durations, likely owing to the two sources 3C120 and 3C454.3 with shorter flare durations compared to the obtained best-fit PSD timescales. For the separations, the correlation with the PSD timescale is strong but does not follow the one-to-one correspondence. Overall, for five of the constrained sources, the mean flare duration matches the PSD timescale better than the flare separation. For four of the sources, both timescales are very close to each other. \cite{mukherjee2019accretion} show that if both separations between the flares and the electron cooling times (resulting in longer flare decays) are of a similar magnitude, they may both contribute to the bend causing a smoother turnover. The constrained sources are also within the 25 sources with the shortest flare durations. Using the flare separations, the scatter is larger. 

\begin{figure}
    \centering
    \includegraphics[width=1\linewidth]{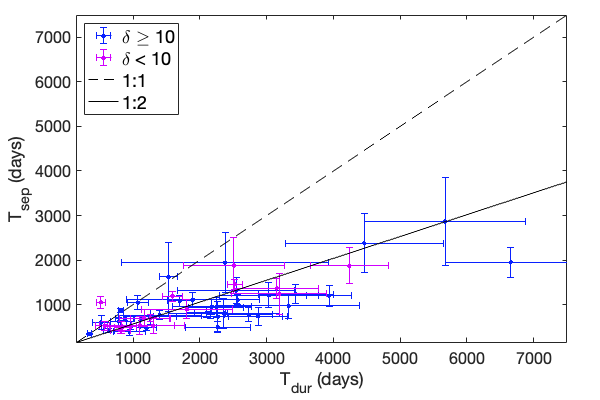}
    \caption{Mean flare separations against the mean flare durations obtained from the decomposition of the light curves using the mean-amplitude limit. Source 3C84 is excluded from the plot for visual reasons, due to a much longer estimated mean flare duration and separation compared to the other sources.}
    \label{fig:TdurTsepeq}
\end{figure}

The flare durations are, on average, twice the length of the separations. However, for the 11 constrained sources this ratio is smaller with \(\sim\)1.3. Figure \ref{fig:TdurTsepeq} shows the flare separations plotted against the flare durations for each light curve. The solid line is the equivalence of twice the duration compared to the separation, whereas the dashed line gives the one-to-one equivalence. We obtained a strong positive correlation, with \(\tau \approx 0.58\) and \(p \approx 7.9\cdot10^{-10}\). Given the longer values obtained for the flare durations, it seems likely that the PSD bend constrained in Paper I is more closely related to the durations. Even if there was a second bend at higher frequencies, we expect the bending power law used in Paper I to rather estimate the low-frequency bend, where the flattening to white noise is expected. However, it is also possible that the flattening observed in some of the sources is not to white noise but to a slope of 1, in which case the flare durations could be probing the high-frequency bend instead. This is potentially the case with 3C120; The best-fit PSD bend of 3C120 is at 2500 days, but its mean flare duration and separation are only $\sim$500 and $\sim$1000 days, respectively. If only the very brightest flares are considered, that is, those with amplitudes over 50 \% of the maximum amplitude in a given posterior draw, the mean separation is \(\sim\) 2600 days. It is then possible that using a two-bend model for the PSD of 3C120 could reveal a high-frequency bend better associated with the durations. We leave this analysis for future studies.

\subsubsection{Connection with the central engine}

\begin{figure*}
\sidecaption
    \includegraphics[width=12cm]{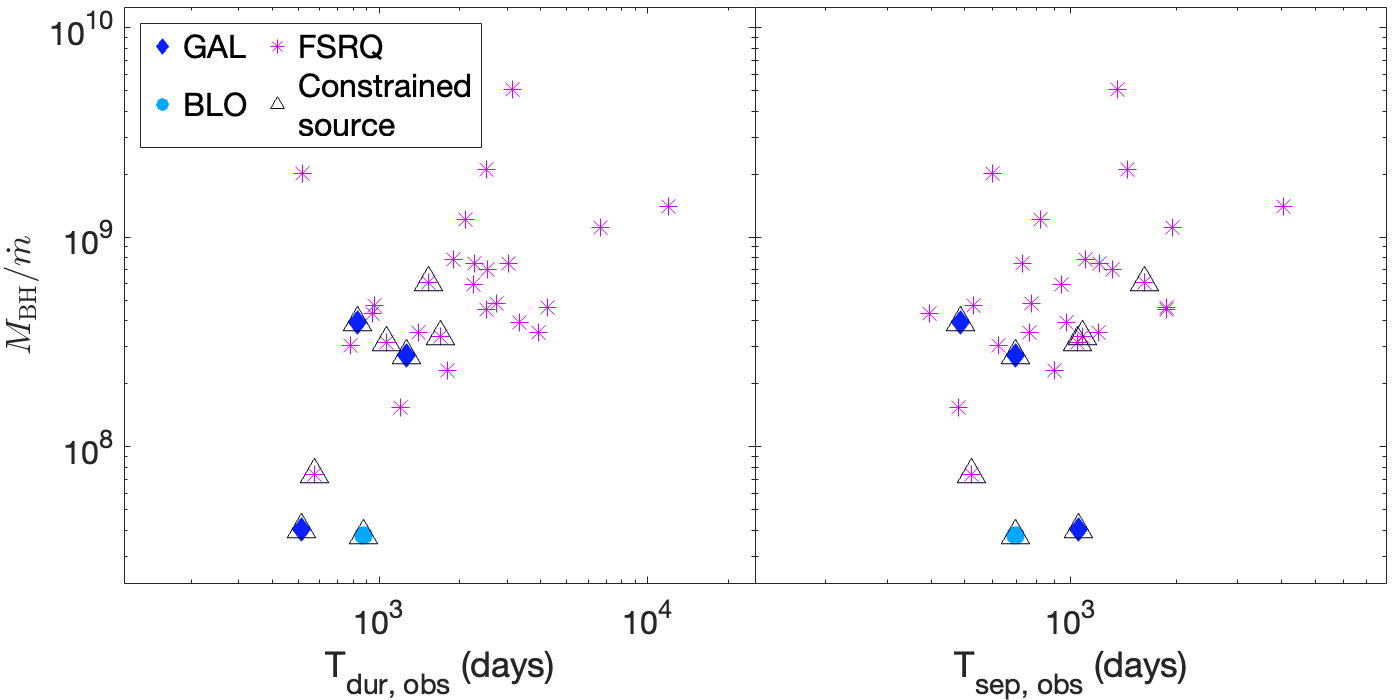}
    \caption{Black hole mass divided by the normalised mass accretion rate ($M_{\rm BH}/\dot m$) against the observed mean flare duration (left) and separation (right). }
    \label{fig:Mbhmandtdur}
\end{figure*}

\begin{table} 
\renewcommand{\arraystretch}{1.3}
\caption{Results from the Kendall rank correlation analysis for associations between \(M_{\rm BH}/\dot m\) and flare durations and separations.}           
\centering          
\begin{tabular}{c c c c l l l }    
\hline\hline   
\(M_{\rm BH}/\dot m\) & \(\tau\) & \(p\) &\\ 
\hline                    
   \(T_{\rm dur, obs}\) & 0.398 & 0.002 \\
   \(T_{\rm sep, obs}\) & 0.310 & 0.016 \\
   \(T_{\rm dur,em}\) & 0.260 & 0.045 \\
   \(T_{\rm sep,em}\) & 0.126 & 0.338 \\
\hline           
\end{tabular}
\label{table:PSDCE}
\tablefoot{\(T_{\rm dur, obs}\) are the observed mean flare durations and \(T_{\rm dur, em}\) the flare durations corrected for both redshift and Doppler boosting. \(T_{\rm sep, obs}\) are the observed mean flare separations and \(T_{\rm sep, em}\) the flare separations corrected for redshift}.
\end{table}

Due to the found connection between the light-curve decomposition results and the PSD timescales, we used both the mean flare durations and separations as potential proxies for the PSD timescale. This is because the number of constrained sources (11) is too low for the analysis and is reduced to only eight for the ones with a black hole mass estimate. Motivated by \cite{mchardy2006active}, who found the PSD timescales of AGNs in the X-ray domain to scale as \(\approx M_{\rm BH}^{1.12}/\dot m^{0.98}\), we estimated the PSD timescale association with \({M_{\rm BH}/\dot m}\) using the black hole masses in units of solar mass (see Fig. 4).

With a sample of 30 sources, we find that both the flare durations and separations have a moderate positive correlation with \(M_{\rm BH}/\dot m\) (Table \ref{table:PSDCE}). The correlations may be driven by the black hole mass, as no correlations exist with the mass accretion rate \(\dot m\) and either timescale. However, for the flare separations, the correlation with \(M_{\rm BH}\) is marginally weakened (\(\tau\) = 0.254, \(p\)= 0.052) compared to \(M_{\rm BH}/\dot m\). Thus, we cannot definitively exclude the need for the division by \(\dot m\) and a larger sample of sources or more accurate estimates for the black hole mass and accretion rate are required to analyse the connection further. 

If the flare durations and separations are corrected to the emission frame, the correlation is weakened for the flare durations and for the flare separations we cannot reject the null hypothesis of no association (Table \ref{table:PSDCE}). The implications are discussed in Sect. \ref{PSDdiscussion}.
 
\subsection{Jet and the central engine}

\begin{table} \label{flaredurcorr}
\renewcommand{\arraystretch}{1.3}
\caption{Results from the Kendall rank correlation analysis for associations between flare rise time and jet and central engine parameters.}         
\centering          
\begin{tabular}{l c c l cc}    
\hline\hline       
\(\tau_{\rm rise, obs}\) & \(\tau\) & \(p\) & \(\tau_{\rm rise, em}\) & \(\tau\) & \(p\)\\ 
\hline                    
   \(M_{\rm BH}\) & 0.291 & 0.026 & \(M_{\rm BH}\) & 0.235 & 0.071 \\
   \(L_{\rm acc}\) & 0.211 & 0.083 & \(L_{\rm acc}\) & 0.260 & 0.032 \\
   \(\dot m\) & 0.131 & 0.321 & \(\dot m\) & 0.214 & 0.101  \\
   \(\Gamma\) & 0.058 & 0.626 & \(\Gamma\) & 0.211 &  0.030   \\
\hline           
\end{tabular}
\label{table:risevscentraleng}
\end{table}

\begin{figure*}
    \sidecaption
    \includegraphics[width=12cm]{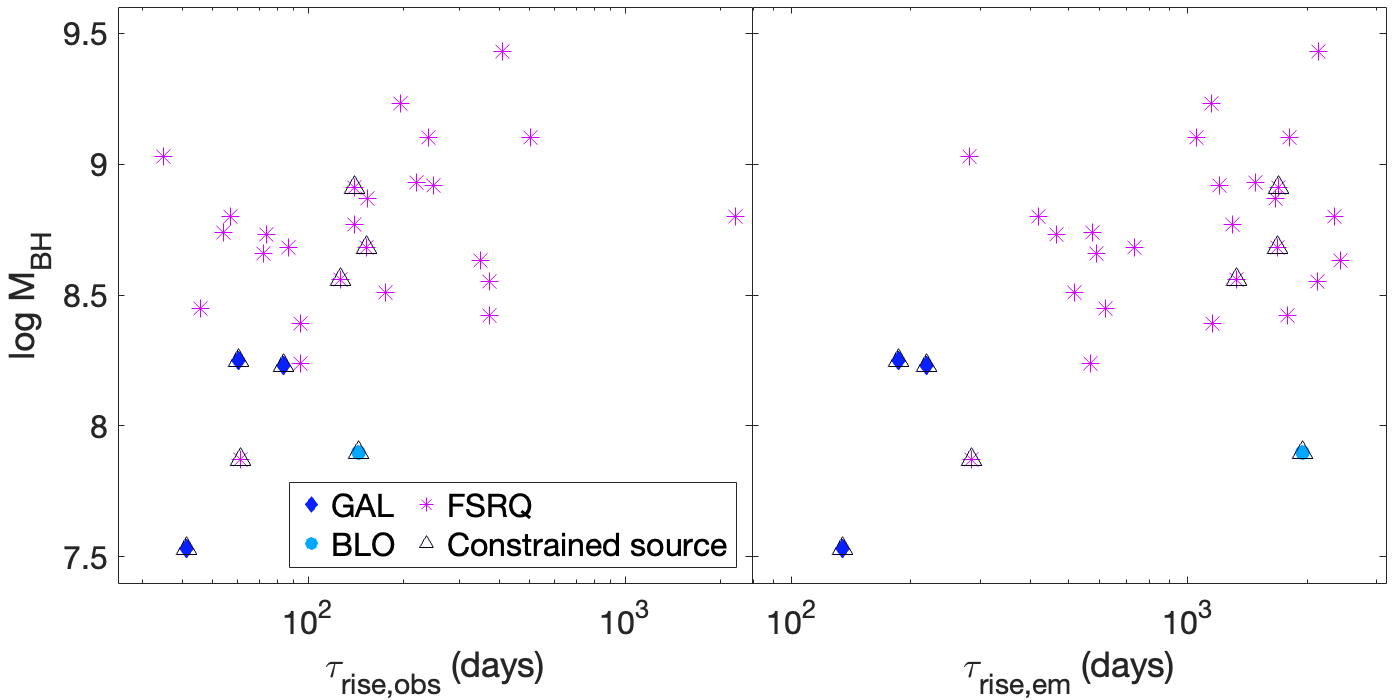}
    \caption{Black hole mass against the median flare rise time in the observed frame (left) and the emission frame (right).}
    \label{fig:MBH}
\end{figure*}

We analysed the data for an association between the black hole mass and the width of the transition region approximated from the median rise time, \(\tau_{\rm rise}\), of the most dominant flares. Fig. \ref{fig:MBH} shows the black hole mass against the rise time in both the observed frame \(\tau_{\rm rise,obs}\) and the redshift and Doppler-corrected emission frame \(\tau_{\rm rise,em}\). We applied the Kendall rank correlation analysis to probe for an association between \(M_{\rm BH}\) and \(\tau_{\rm rise,em}\) and were not able to reject the null hypothesis (Table \ref{table:risevscentraleng}). However, if the source 0235+164 with weak emission lines and thus uncertain black hole mass is removed from the analysis, a significant weak positive correlation, with \(\tau \approx 0.29\) and \(p \approx 0.03\), is obtained. A similar correlation, including all of the sources, is obtained for \(\tau_{\rm rise,obs}\) and \(M_{\rm BH}\).

\begin{figure*}
    \sidecaption
    \includegraphics[width=12cm]{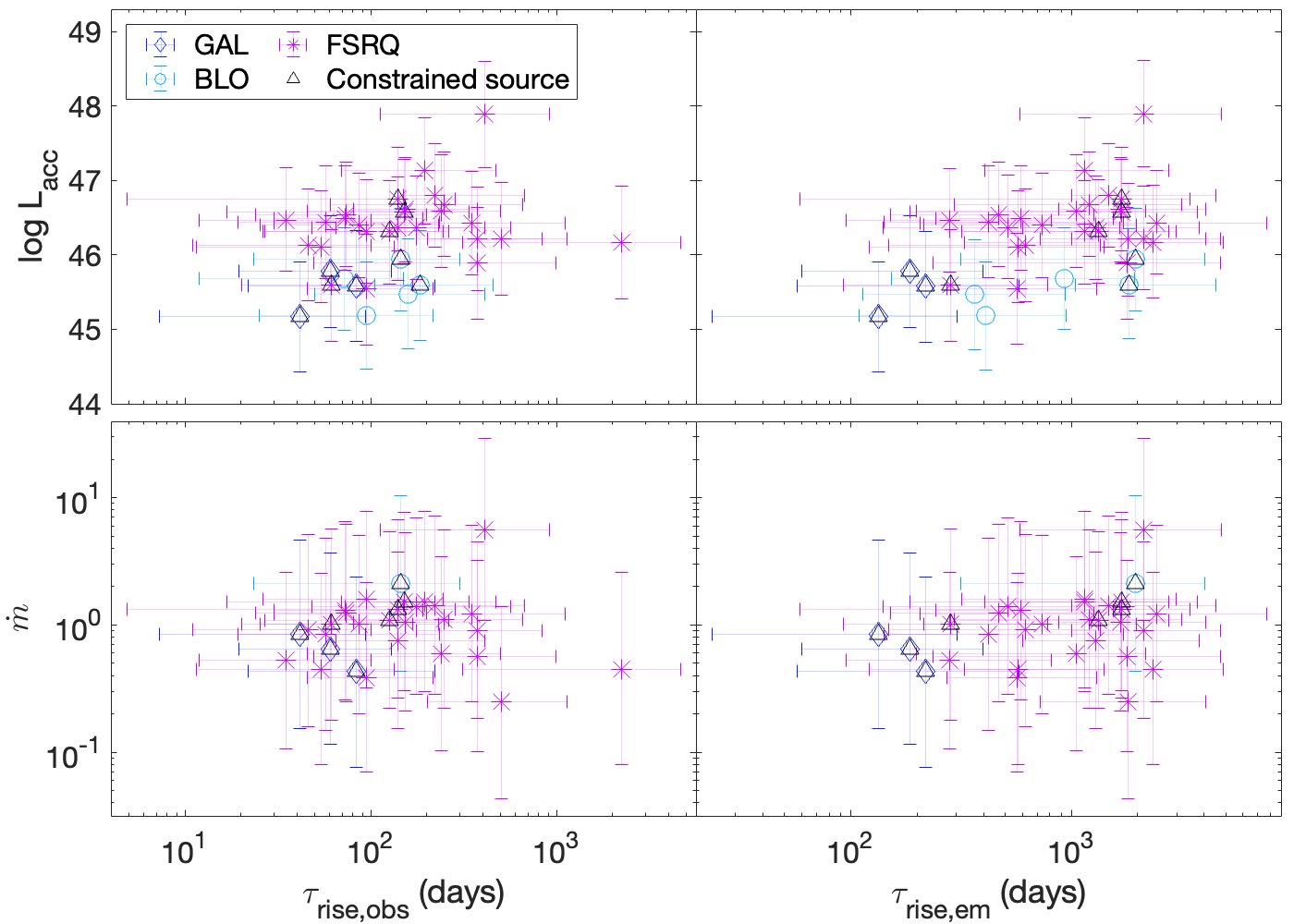}
    \caption{Accretion disk luminosity (top row) and normalised mass accretion rate (bottom row) against the observed median flare rise time (left) and emission-frame rise time (right).}
    \label{fig:MA_TRise}
\end{figure*}

As in the previous analysis, we used \(\tau_{\rm rise,em}\) as a proxy for the emission-region size, and we calculated the accretion disk luminosity for 34 sources with estimated Doppler factors. Figure \ref{fig:MA_TRise} shows in the top row the accretion disk luminosity \(L_{\rm acc}\) against both the observed and the emission-frame rise times. We found a weak significant positive correlation between \(L_{\rm acc}\) and \(\tau_{\rm rise,em}\) (Table \ref{table:risevscentraleng}). The result is the same as for the mass accretion rate \(\dot M\), when using the same radiative efficiency for all sources. We calculated the normalised accretion rate \(\dot m\) using the accretion disk luminosity and the calculated black hole masses for 30 sources. Figure \ref{fig:MA_TRise} shows \(\dot m\) against the observed and emission-frame rise times. Here, we could not reject the null hypothesis of no correlation. Neither parameter has a significant correlation with \(\tau_{\rm rise,obs}\).

\begin{figure*}
    \sidecaption
    \includegraphics[width=12cm]{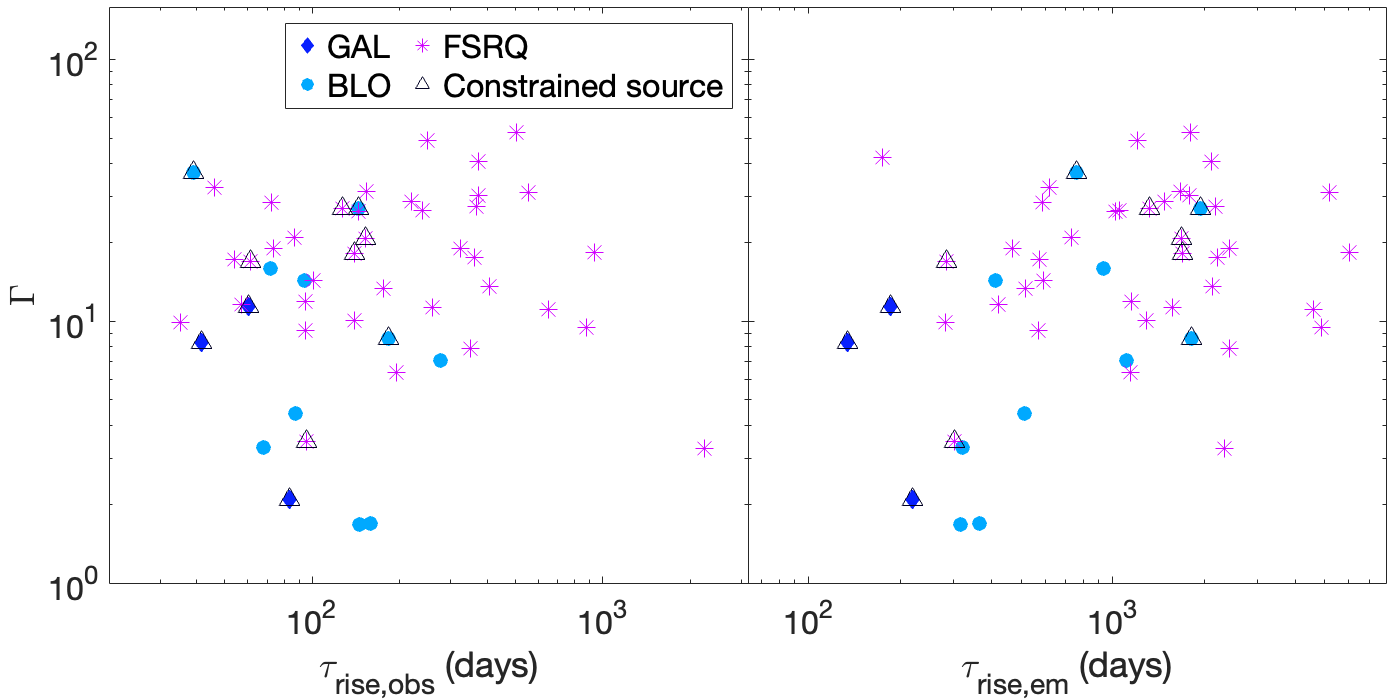}
    \caption{Lorentz factor against the median rise time in the observed frame (left) and the emission frame (right).}
    \label{fig:LorentzvsRise}
\end{figure*}

With the 51 sources for which we were able to determine the Lorentz factor, and testing for an association with the emission-frame rise time \(\tau_{\rm rise,em}\), we found a weak positive correlation (Table \ref{table:risevscentraleng}). No association was found with \(\tau_{\rm rise,obs}\). Figure \ref{fig:LorentzvsRise} shows the bulk jet Lorentz factor \(\Gamma\) against \(\tau_{\rm rise,obs}\) and against \(\tau_{\rm rise,em}\). 

\section{Discussion} \label{Discussion}

We have analysed the decomposed light curves to understand which timescales could correspond to the PSD timescales found in Paper I and whether they can be connected to the black hole mass and accretion rate. We have also used a simplified assumption that the derived rise times can be used as a proxy for the emission region size of the jet and that it corresponds to flares observed to begin at the transition region. Here, we discuss some of the implications of the potential correlations as well as the caveats in the analysis. Because the number of BLOs and especially GALs is small in the sample, we cannot make strong inferences between the source types. 

\subsection{Associations with the emission region size}

\cite{kovalev2020transition} analysed VLBI data to look for jet shape transitions from parabolic to conical. They found the transition region in ten sources, three of which are included in our sample: 0415+379, 3C120, and BL Lac. Comparing the emission-frame rise times of the three sources to the jet widths given in \cite{kovalev2020transition}, the width of the transition region in parsecs agrees in ordering with the rise times obtained from the light-curve decomposition: In our analysis, BL Lac has the longest emission-frame rise time with 411 days, 0415+379 with 186 days, and 3C120 with 134 days. We note that the estimated rise times in our analysis are medians, and we use the e-folding timescale instead of a double-folding timescale. While the number of sources is too low to make inferences, this result is in agreement with our simplified assumption that the rise times can be used as a proxy for the width of the transition region.

\subsubsection{Lorentz factor} \label{Lorentzdiscussion}

Previous studies have suggested that the jet bulk Lorentz factor of AGNs may be higher for sources with longer acceleration regions (e.g. \citealt{vlahakis2003relativistic}; \citealt{beskin2006effective}; \citealt{lyubarsky2009asymptotic}; \citealt{kutkin2019opacity}). Consequently, \cite{potter2015new} showed that a larger width of the transition region is connected to higher Lorentz factors. A related conclusion was obtained by \cite{malzac2018jet}, who analysed the behaviour of a BHXRB with a shell-shock model and found that a higher average jet Lorentz factor caused shocks to be formed further downstream of the jet increasing the observed timescales. Their analysis did not consider any specific emission regions in relation to the Lorentz factor, but the conclusion of larger jet Lorentz factors connected to larger jet radii may suggest a similar effect on AGN observations. While we found an association between the Lorentz factor and the rise times, the strength of the correlation is low. This could be in part due to different scaling factors between the length of the accelerating region and the width of the transition region for different sources; \cite{kovalev2020transition} also estimated the deprojected break locations from the central black hole and found 0415+379 to have a longer parabolic region compared to BL Lac. This may then indicate a more complex relationship between the length of the accelerating region and the width of the transition region.

\subsubsection{Black hole mass} \label{blackholemass}

We found a potential weak positive correlation between the rise times and the black hole masses. Assuming a simple relationship with the length of the accelerating region and width at the jet shape transition, we may expect to observe a relationship. This is because some studies have found the location of this transition to coincide with the Bondi radius or the sphere of gravitational influence (SGI) (e.g. \citealt{asada2012structure}; \citealt{tseng2016structural}; \citealt{kovalev2020transition}). However, this relation is likely complicated: \cite{okino2022collimation} analysed the location of the jet collimation break in 3C273 and compared it to the jet shape transition regions obtained for other sources in relation to the location of the SGI. For 3C273, they found the transition in the jet shape to be downstream of the SGI, and the other sources had a large scatter in the locations of the jet shape transitions around their respective SGIs. \cite{okino2022collimation} note that this scatter in the locations indicates that the surrounding environment contributes to the location of the transition. \cite{boccardi2021jet} suggest that an upstream location of the jet break (from the Bondi radius) could occur especially with low-luminosity sources, and \cite{fariyanto2025jet} discuss how this could be connected to a lower accretion rate, which may also affect the collimation profile of the jet.

\subsubsection{Accretion disk luminosity}

We also found a weak significant positive correlation between the optical accretion disk luminosity (and thus un-normalised mass accretion rate) and flare rise times. In X-ray variability studies of AGNs, it has been shown that a higher X-ray luminosity corresponds to a longer variability timescale (e.g. \citealt{barr1986limits}; \citealt{lawrence1993x}; \citealt{green1993nature}). \cite{barr1986limits} compared the doubling times of the intensity of the X-ray emission to the 2-10 keV X-ray luminosity, finding a clear positive correlation. Similarly, \cite{hovatta2007statistical} found that higher radio luminosities lead to longer variability timescales. \cite{sbarrato2014jet} analysed a sample of blazars and GALs and found the BLR luminosity to be closely correlated with radio luminosity. Thus, our results of the connection between accretion disk luminosity and flare rise times are not surprising, given the correlation found in \cite{hovatta2007statistical} and may point to a common dependency between source luminosity and variability timescales across the electromagnetic spectrum.

\begin{table} \label{flaredurcorr}
\renewcommand{\arraystretch}{1.3}
\caption{Results from the Kendall rank correlation analysis for associations between redshift and emission-frame rise time, Lorentz factor, and central engine parameters.}       
    
\centering          
\begin{tabular}{l c c c l l l }    
\hline\hline       
                     
\(z\) & \(\tau\) & \(p\) \\ 
\hline                    
   \(\tau_{\rm rise,em}\) & 0.369 &  \(1.177\cdot10^{-4}\)   \\
   \(M_{\rm BH}\) & 0.446 & \(6.088\cdot10^{-4}\) \\
   \(L_{\rm acc}\) & 0.634 & \(1.529\cdot10^{-7}\) \\
   \(\dot m\) & 0.318 & 0.015  \\
   \(\Gamma\) & 0.184 & 0.058  \\

\hline           
\end{tabular}
\label{table:redshift}
\end{table}

\subsubsection{Redshift}

The found correlations are mostly weak, and their significances in either direction are typically driven by a small number of sources. We examined whether the correlations between the rise time and the jet and central engine parameters could be driven by a common dependency on redshift. The results from the correlation analysis are given in Table \ref{table:redshift}. The positive correlations between redshift and emission-frame rise time, black hole mass, accretion disk luminosity, and normalised mass accretion rate are significant. Thus, we performed a partial correlation analysis using the ppcor R-package \citep{kim2015ppcor} to examine whether redshift explains the correlation between the emission-frame rise time and black hole mass and accretion disk luminosity. Indeed, neither correlation is significant when considering redshift as a third variable (\(\tau\) = 0.126, \(p\) = 0.337 and \(\tau\) = 0.075, \(p\) = 0.541, respectively),  and we can conclude that the original correlations are driven by redshift. While the Lorentz factor correlation with redshift is not significant, the margin to the significance limit is very small. If a partial correlation accounting for redshift is performed between the Lorentz factor and the rise time, the association weakens to \(\tau\) = 0.156 and \(p\) = 0.110.

\subsection{PSD timescales}\label{PSDdiscussion}

We found that both the observer-frame flare separations and durations correlate with \(M_{\rm BH}/\dot m\), which is potentially driven by a correlation with \(M_{\rm BH}\). The association between the observer-frame rise times and black hole masses is then likely related to the correlation with the flare durations. However, if only FSRQs are considered in the analysis, a statistically significant association persists only between the timescales and \(M_{\rm BH}/\dot m\). Further analyses are thus required to understand the connection better.

The correlation with the redshift-corrected flare separations and \(M_{\rm BH}/\dot m\) did not persist. This is surprising as the redshifts have a relatively narrow range of values and the resulting corrections to the timescales are moderate; a correlation analysis between \(M_{\rm BH}/\dot m\) and only redshift-corrected flare durations gives a similar result as the analysis between \(M_{\rm BH}/\dot m\) and observer-frame flare durations. Thus, we examined whether a common association with redshift may affect the results. Indeed, the Doppler- and redshift-corrected flare durations have a positive association with redshift whereas the redshift-corrected flare separations have a negative correlation with redshift (Table \ref{table:redshiftpsd}).

While the positive association between \(M_{\rm BH}/\dot m\) and redshift is not significant (\(\tau\) = 0.230, \(p\) = 0.077), it is possible that this association then affects the analysis. Using redshift as a third variable, we find that the correlation between \(M_{\rm BH}/\dot m\) and the Doppler- and redshift-corrected flare durations decreases to \(\tau\) = 0.204 and \(p\) = 0.120, whereas with the redshift-corrected flare separations the correlation is increased to \(\tau\) = 0.222 and \(p\) = 0.090. Nevertheless, we cannot reject the null hypothesis of no correlation. 

As \cite{mchardy2009x} suggested, a connection with the X-ray PSD bend of a blazar implies that a central-engine source modulates the variations in the jet. This is especially understandable with the flare separations as the rate of ejections of superluminal knots could be related to the central engine properties and matter infall (\citealt{marscher2002observational}; \citealt{chatterjee2009disk}; \citealt{chatterjee2011connection}). On the other hand, the association with the flare durations and the black hole mass could be partially explained by the emission region size being further out from the black hole, as explained in Sect. \ref{blackholemass}. The mass accretion rate could then also be related to the location of the emitting region (e.g. \citealt{fariyanto2025jet}).

While we cannot verify the emission-frame timescale association, the results are intriguing as they suggest that the bend seen in the PSDs of radio light curves may have a connection to the central engine, similarly as in the X-ray domain.

\subsubsection{Connection between flare durations and separations} \label{redshiftconn}
            
\begin{figure}
    \centering
    \includegraphics[width=1\linewidth]{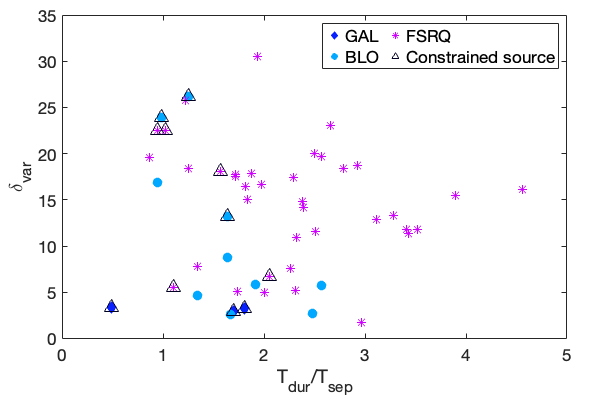}
    \caption{Ratio of the mean flare durations and separations against the estimated variability Doppler factor. The source 0836+710, with a very high Doppler factor of \(\sim 79\), is excluded from the plot.}
    \label{fig:ratiovsdoppler}
\end{figure}

\begin{table} \label{flaredurcorr}
\renewcommand{\arraystretch}{1.3}
\caption{Results from the Kendall rank correlation analysis for associations between redshift and emission-frame flare durations and separations.}             
\centering          
\begin{tabular}{c c c c l l l }    
\hline\hline          
\(z\) & \(\tau\) & \(p\) \\ 
\hline                    
   \(T_{\rm dur,em}\) & 0.410 &  \(1.894\cdot10^{-5}\)   \\
   \(T_{\rm sep,em}\) & -0.355 & \(1.552\cdot10^{-4}\) \\
\hline           
\end{tabular}
\label{table:redshiftpsd}
\end{table}

\cite{lister2001relativistic} hypothesised that, assuming similarity of sources, higher Doppler-boosting factors would lead to shorter observed flare durations compared to their separations. However, we find no correlation between the Doppler factor and \(T_{\rm dur,obs}/T_{\rm sep,obs}\) (Fig. \ref{fig:ratiovsdoppler}). One potential explanation is that the flare durations are longer in the emission frame for sources with higher Lorentz factors, related to the increased width of the transition region \citep{potter2015new}. This could then naturally counteract the otherwise increasing ratio between the flare durations and separations. Of course, this interpretation assumes that the rate of ejections is not significantly different between sources. 

As mentioned above, the results in Table \ref{table:redshiftpsd} suggest that the intrinsic flare durations are longer for the more distant sources, and that the intrinsic rate of ejections increases (flare separations decrease) with a higher redshift. The former association is driven by BLOs and GALs, as the association weakens and is no longer significant with FSRQs alone. The negative correlation between the intrinsic rate of ejections and redshift is for both the two source divisions (FSRQs and BLOs/GALs) and the entire sample. The division to BLOs/GALs is due to their generally smaller redshifts in the sample and because the number of sources is too low to isolate them into two groups. The increasing flare durations with redshift may also be a sample bias as we have chosen sources that have discernible flares. \cite{lister2001relativistic} notes that due to severe overalapping of flares in low-beamed sources, they will typically have fairly stable observed flux densities.

Because of this opposite correlation between the emission-frame flare durations and redshift and redshift-corrected flare separations and redshift, the two correlations could counteract each other and explain why we do not see longer observed flare separations with shorter observed flare durations for the more Doppler-boosted sources, which generally are at higher redshifts. On the other hand, if redshift is accounted for through partial correlation, the emission-frame flare durations and redshift-corrected flare separations also correlate weakly ($\mathrm{\tau \approx  0.29, p \approx 3\cdot10^{-3}}$). 

\subsubsection{VLBI timescales}

In Paper I, we found a potential association between the PSD timescale and the longest time a 43 GHz VLBI knot was observed for, and we hypothesised that the timescale may be related to the flare durations. This association is now strengthened with the good correspondence between the flare durations and the PSD timescales. However, in general, the VLBI knot ejections have been associated with the decay times of the flares rather than their durations (e.g. \citealt{turler1999modelling}; \citealt{lindfors2006synchrotron}). This is because the flare is expected to peak at the radio core, only after which an associated knot can be observed to separate from the core. We do not have an answer for this apparent conflict; it is possible that these most-dominant flares are low-peaking ones \citep{valtaoja1992five} in which case the shock becomes optically thin already before the peak amplitude. It is then related to the resolution of the VLBI network and the distance of the source, which determine when the knot is observed. On the other hand, how long the knot is visible in the VLBI image also depends on the sensitivity of the array and the ability to model it as a separate feature, and therefore it may be considered a part of the flare in the 37 GHz single-dish data for a longer time.

Another explanation is that the preliminary correspondence between the PSD and VLBI knot timescale is only at an approximately correct value: In this analysis, we did not isolate and measure the absolute longest high-amplitude flares, but included all of the brightest flares to approximate the most dominant flare durations and separations. Thus, it is possible that the VLBI knot and PSD timescales also correspond to the decay time of a single dominant bright flare and it is approximately of a similar value as the combined durations of a selection of the brightest flares.

\section{Conclusions}

We have analysed 54 sources to compare the decomposed flare durations and separations either to the constrained PSD timescales or to the unconstrained estimates obtained in Paper I. We found the strongest one-to-one association between the PSD timescale and the mean duration of the brightest flares, whereas the flare separations also correlate with both the PSD timescale and the flare durations. However, as only 11 of the sources in the sample have a constrained PSD timescale, we cannot make definitive statements on whether the durations or separations are more closely associated with the PSD bend. 

A connection with the PSD timescale and \(M_{\rm BH}/\dot m\) was found through using the flare durations and separations as proxies for the PSD timescale. While both of the decomposed timescales had a significant association with \(M_{\rm BH}/\dot m\) when using observer-frame values, the association was no longer significant in the emission frame. Further analyses are thus required to understand the connection better and whether the accuracy of the estimated central engine parameters affects the results.

We also found that both the emission-frame flare durations and separations correlate with redshift, with both the flare durations and the rate of flares increasing for higher redshift sources. Curiously, if redshift is accounted for, longer intrinsic flare durations are associated with longer intrinsic flare separations (lower rate of flares). It is unclear, whether this association is then driven by differences in the sources or whether these connections are induced by, for example, a sample bias. 

Using the flare rise times as a proxy for the emission region size, we found weak correlations between the emission-frame flare rise times and the jet and central engine parameters. However, the associations appear to be driven by a common dependency on redshift. This does not necessarily mean that no correlation exists, as the analysis may be limited by the sample being both relatively small and including only bright sources at higher redshifts.

\begin{acknowledgements}
The authors would like to thank Dr. Yannis Liodakis and Dr. Elina Lindfors for their valuable help when preparing the manuscript. S.K. was supported by Emil Aaltonen Foundation and Jenny and Antti Wihuri Foundation. T.H. acknowledges support from the Research Council of Finland projects 317383, 320085, 345899, and 362571 and from the European Union ERC-2024-COG - PARTICLES - 101169986. Views and opinions expressed are however those of the author(s) only and do not necessarily reflect those of the European Union or the European Research Council Executive Agency. Neither the European Union nor the granting authority can be held responsible for them. This publication makes use of data obtained at the Metsähovi Radio Observatory, operated by the Aalto University ({https://www.metsahovi.fi/opendata/}). This research has made use of data from the MOJAVE database that is maintained by the MOJAVE team \citep{lister2018mojave}.
\end{acknowledgements}

\bibliographystyle{aa} 
\bibliography{sources} 

@article{huppenkothen2015dissecting,
  title={Dissecting magnetar variability with Bayesian hierarchical models},
  author={Huppenkothen, Daniela and Brewer, Brendon J and Hogg, David W and Murray, Iain and Frean, Marcus and Elenbaas, Chris and Watts, Anna L and Levin, Yuri and Van Der Horst, Alexander J and Kouveliotou, Chryssa},
  journal={ApJ},
  volume={810},
  number={1},
  pages={66},
  year={2015},
  publisher={IOP Publishing}
}

@article{aghanim2020planck,
  title={Planck 2018 results. VI. Cosmological parameters},
  author={{Planck Collaboration} and {Aghanim}, N. and {Akrami}, Y. and {Ashdown}, M. and {Aumont}, J. and {Baccigalupi}, C. and {Ballardini}, M. and {Banday}, A.~J. and {Barreiro}, R.~B. and {Bartolo}, N. and {Basak} and others},
  journal={A\&A},
  volume={641},
  pages={A6},
  year={2020}
}

@article{asada2012structure,
  title={The structure of the M87 jet: a transition from parabolic to conical streamlines},
  author={Asada, Keiichi and Nakamura, Masanori},
  journal={ApJL},
  volume={745},
  number={2},
  pages={L28},
  year={2012},
  publisher={IOP Publishing}
}

@article{barr1986limits,
  title={Limits of X-ray variability in active galactic nuclei},
  author={Barr, P and Mushotzky, RF},
  journal={Nature},
  volume={320},
  number={6061},
  pages={421--423},
  year={1986},
  publisher={Nature Publishing Group UK London}
}

@article{belloni1990variability,
  title={Variability in the noise properties of Cygnus X-1},
  author={Belloni, Tomaso and Hasinger, G{\"u}nther},
  journal={A\&A},
  volume={227},
  pages={L33--L36},
  year={1990}
}

@article{brewer2018dnest4,
  title={DNest4: Diffusive nested sampling in C++ and Python},
  author={Brewer, Brendon J and Foreman-Mackey, Daniel},
  journal={J. Stat. Softw},
  volume={86},
  pages={1--33},
  year={2018}
}

@article{brewer2011diffusive,
  title={Diffusive nested sampling},
  author={Brewer, Brendon J and P{\'a}rtay, Livia B and Cs{\'a}nyi, G{\'a}bor},
  journal={Stat. Comput.},
  volume={21},
  number={4},
  pages={649--656},
  year={2011},
  publisher={Springer}
}

@article{boccardi2021jet,
  title={Jet collimation in NGC 315 and other nearby AGN},
  author={Boccardi, B and Perucho, M and Casadio, C and Grandi, Paola and Macconi, D and Torresi, ELEONORA and Pellegrini, S and Krichbaum, TP and Kadler, M and Giovannini, Gabriele and others},
  journal={A\&A},
  volume={647},
  pages={A67},
  year={2021},
  publisher={EDP Sciences}
}

@article{beskin2006effective,
  title={The effective acceleration of plasma outflow in the paraboloidal magnetic field},
  author={Beskin, VS and Nokhrina, EE},
  journal={MNRAS},
  volume={367},
  number={1},
  pages={375--386},
  year={2006},
  publisher={Blackwell Science Ltd 23 Ainslie Place, Edinburgh EH3 6AJ, UK. Telephone~…}
}

@book{bottcher2012relativistic,
  title={Relativistic jets from active galactic nuclei},
  author={B{\"o}ttcher, Markus and Harris, Daniel E and Krawczynski, Henric},
  year={2012},
  publisher={John Wiley \& Sons}
}

@article{cohen1987nature,
  title={The nature of the BL Lacertae object AO 0235+ 164},
  author={Cohen, Ross D and Smith, Harding E and Junkkarinen, Vesa T and Burbidge, E Margaret},
  journal={ApJ},
  volume={318},
  pages={577--584},
  year={1987}
}

@article{chatterjee2009disk,
  title={DISK--JET CONNECTION IN THE RADIO GALAXY 3C 120},
  author={Chatterjee, Ritaban and Marscher, Alan P and Jorstad, Svetlana G and Olmstead, Alice R and McHardy, Ian M and Aller, Margo F and Aller, Hugh D and L{\"a}hteenm{\"a}ki, Anne and Tornikoski, Merja and Hovatta, Talvikki and others},
  journal={ApJ},
  volume={704},
  number={2},
  pages={1689},
  year={2009},
  publisher={IOP Publishing}
}

@article{chatterjee2011connection,
  title={Connection between the accretion disk and jet in the radio galaxy 3C 111},
  author={Chatterjee, Ritaban and Marscher, Alan P and Jorstad, Svetlana G and Markowitz, Alex and Rivers, Elizabeth and Rothschild, Richard E and McHardy, Ian M and Aller, Margo F and Aller, Hugh D and L{\"a}hteenm{\"a}ki, Anne and others},
  journal={ApJ},
  volume={734},
  number={1},
  pages={43},
  year={2011},
  publisher={IOP Publishing}
}

@article{daly1988gasdynamics,
  title={The gasdynamics of compact relativistic jets},
  author={Daly, Ruth A and Marscher, Alan P},
  journal={ApJ},
  volume={334},
  pages={539--551},
  year={1988}
}

@article{fariyanto2025jet,
  title={Jet Collimation Profile of the Low-luminosity Active Galactic Nucleus M84: Insight into the Jet Formation in the Low-accretion Regime},
  author={Fariyanto, Elika P and Hada, Kazuhiro and Cui, Yuzhu and Honma, Mareki and Nakamura, Masanori and Asada, Keiichi and Wang, Xuezheng and Jiang, Wu},
  journal={ApJ},
  volume={991},
  number={1},
  pages={13},
  year={2025},
  publisher={IOP Publishing}
}

@article{fender2004towards,
  title={Towards a unified model for black hole X-ray binary jets},
  author={Fender, Rob P and Belloni, T Mꎬ and Gallo, Elena},
  journal={MNRAS},
  volume={355},
  number={4},
  pages={1105--1118},
  year={2004},
  publisher={Blackwell Science Ltd}
}

@article{fromm2015location,
  title={On the location of the supermassive black hole in CTA 102},
  author={Fromm, Christian M and Perucho, Manel and Ros, Eduardo and Savolainen, Tuomas and Zensus, J Anton},
  journal={A\&A},
  volume={576},
  pages={A43},
  year={2015},
  publisher={EDP Sciences}
}

@article{green1993nature,
  title={On the nature of rapid X-ray variability in active galactic nuclei},
  author={Green, Andrew Robert and McHardy, IM and Lehto, HJ},
  journal={MNRAS},
  volume={265},
  number={3},
  pages={664--680},
  year={1993},
  publisher={Oxford University Press Oxford, UK}
}

@article{gonzalez2012x,
  title={X-ray variability of 104 active galactic nuclei-XMM-Newton power-spectrum density profiles},
  author={Gonzalez-Martin, Omaira and Vaughan, Simon},
  journal={A\&A},
  volume={544},
  pages={A80},
  year={2012},
  publisher={EDP Sciences}
}

@article{hughes1992university,
  title={The University of Michigan radio astronomy data base. I-Structure function analysis and the relation between BL Lacertae objects and quasi-stellar objects},
  author={Hughes, PA and Aller, HD and Aller, MF},
  journal={ApJ},
  volume={396},
  pages={469--486},
  year={1992}
}

@article{kankkunen2025long,
  title={Long-term radio variability of active galactic nuclei at 37 GHz},
  author={Kankkunen, Sofia and Tornikoski, Merja and Hovatta, Talvikki and L{\"a}hteenm{\"a}ki, Anne},
  journal={A\&A},
  volume={693},
  pages={A318},
  year={2025},
  publisher={EDP Sciences}
}

@article{kelly2009variations,
  title={Are the variations in quasar optical flux driven by thermal fluctuations?},
  author={Kelly, Brandon C and Bechtold, Jill and Siemiginowska, Aneta},
  journal={ApJ},
  volume={698},
  number={1},
  pages={895},
  year={2009},
  publisher={IOP Publishing}
}

@article{kim2015ppcor,
  title={ppcor: an R package for a fast calculation to semi-partial correlation coefficients},
  author={Kim, Seongho},
  journal={Commun. Stat. Appl. Methods},
  volume={22},
  number={6},
  pages={665},
  year={2015}
}

@article{koljonen2024galaxy,
  title={Galaxy group-associated distances to very high energy gamma-ray emitting BL Lacs KUV 00311- 1938 and S2 0109+ 22},
  author={Koljonen, Karri II and Lindfors, Elina and Nilsson, Kari and Hein{\"a}m{\"a}ki, Pekka and Kotilainen, Jari},
  journal={MNRAS},
  volume={531},
  number={4},
  pages={5084--5096},
  year={2024},
  publisher={Oxford University Press}
}

@article{kovalev2020transition,
  title={A transition from parabolic to conical shape as a common effect in nearby AGN jets},
  author={Kovalev, Y Y and Pushkarev, A B and Nokhrina, E E and Plavin, A V and Beskin, V S and Chernoglazov, A V and Lister, M L and Savolainen, Tuomas},
  journal={MNRAS},
  volume={495},
  number={4},
  pages={3576--3591},
  year={2020},
  publisher={Oxford University Press}
}

@article{kutkin2014core,
  title={The core shift effect in the blazar 3C 454.3},
  author={Kutkin, AM and Sokolovsky, KV and Lisakov, MM and Kovalev, YY and Savolainen, T and Voytsik, PA and Lobanov, AP and Aller, HD and Aller, MF and Lahteenmaki, A and others},
  journal={MNRAS},
  volume={437},
  number={4},
  pages={3396--3404},
  year={2014},
  publisher={The Royal Astronomical Society}
}

@article{lawrence1993x,
  title={X-ray variability of active galactic nuclei-A universal power spectrum with luminosity-dependent amplitude},
  author={Lawrence, A and Papadakis, I},
  journal={ApJ},
  volume={414},
  pages={L85--L88},
  year={1993}
}

@article{liodakis2018constraining,
  title={Constraining the limiting brightness temperature and Doppler factors for the largest sample of radio-bright blazars},
  author={Liodakis, Ioannis and Hovatta, Talvikki and Huppenkothen, Daniela and Kiehlmann, Sebastian and Max-Moerbeck, Walter and Readhead, Anthony CS},
  journal={ApJ},
  volume={866},
  number={2},
  pages={137},
  year={2018},
  publisher={IOP Publishing}
}

@article{lister2001relativistic,
  title={Relativistic beaming and flux variability in active galactic nuclei},
  author={Lister, Matthew L},
  journal={ApJ},
  volume={561},
  number={2},
  pages={676},
  year={2001},
  publisher={IOP Publishing}
}

@article{lister2018mojave,
  title={MOJAVE. XV. VLBA 15 GHz total intensity and polarization maps of 437 parsec-scale AGN jets from 1996 to 2017},
  author={Lister, ML and Aller, MF and Aller, HD and Hodge, MA and Homan, DC and Kovalev, YY and Pushkarev, AB and Savolainen, T},
  journal={ApJS},
  volume={234},
  number={1},
  pages={12},
  year={2018},
  publisher={IOP Publishing}
}

@article{lister2019mojave,
  title={MOJAVE. XVII. Jet kinematics and parent population properties of relativistically beamed radio-loud blazars},
  author={Lister, ML and Homan, DC and Hovatta, T and Kellermann, KI and Kiehlmann, S and Kovalev, YY and Max-Moerbeck, W and Pushkarev, AB and Readhead, ACS and Ros, E and others},
  journal={ApJ},
  volume={874},
  number={1},
  pages={43},
  year={2019},
  publisher={IOP Publishing}
}

@article{lahteenmaki1999total,
  title={Total flux density variations in extragalactic radio sources. III. Doppler boosting factors, lorentz factors, and viewing angles for active galactic nuclei},
  author={L{\"a}hteenm{\"a}ki, A and Valtaoja, E},
  journal={ApJ},
  volume={521},
  number={2},
  pages={493},
  year={1999},
  publisher={IOP Publishing}
}

@article{lyubarsky2009asymptotic,
  title={Asymptotic structure of Poynting-dominated jets},
  author={Lyubarsky, Yuri},
  journal={ApJ},
  volume={698},
  number={2},
  pages={1570},
  year={2009},
  publisher={IOP Publishing}
}

@article{malzac2018jet,
  title={A jet model for the fast IR variability of the black hole X-ray binary GX 339-4},
  author={Malzac, Julien and Kalamkar, Maithili and Vincentelli, Federico and Vue, Alexis and Drappeau, Samia and Belmont, Renaud and Casella, Piergiorgio and Clavel, Maica and Corbel, Stphane and Coriat, Micka{\"e}l and others},
  journal={MNRAS},
  volume={480},
  number={2},
  pages={2054--2071},
  year={2018},
  publisher={Oxford University Press}
}

@article{marscher2002observational,
  title={Observational evidence for the accretion-disk origin for a radio jet in an active galaxy},
  author={Marscher, Alan P and Jorstad, Svetlana G and G{\'o}mez, Jos{\'e}-Luis and Aller, Margo F and Ter{\"a}sranta, Harri and Lister, Matthew L and Stirling, Alastair M},
  journal={Nature},
  volume={417},
  number={6889},
  pages={625--627},
  year={2002},
  publisher={Nature Publishing Group UK London}
}

@article{mchardy2006active,
  title={Active galactic nuclei as scaled-up Galactic black holes},
  author={McHardy, Ian M and Koerding, E and Knigge, C and Uttley, P and Fender, RP},
  journal={Nature},
  volume={444},
  number={7120},
  pages={730--732},
  year={2006},
  publisher={Nature Publishing Group UK London}
}

@incollection{mchardy2009x,
  title={X-Ray variability of AGN and relationship to Galactic black hole binary systems},
  author={McHardy, Ian},
  booktitle={The Jet Paradigm: From Microquasars to Quasars},
  pages={203--232},
  year={2009},
  publisher={Springer}
}

@article{mukherjee2019accretion,
  title={The accretion disc-jet connection in blazars},
  author={Mukherjee, Sagnick and Mitra, Kaustav and Chatterjee, Ritaban},
  journal={MNRAS},
  volume={486},
  number={2},
  pages={1672--1680},
  year={2019},
  publisher={Oxford University Press}
}

@article{ghisellini2011transition,
  title={The transition between BL Lac objects and flat spectrum radio quasars},
  author={Ghisellini, G and Tavecchio, Foschini and Foschini, L and Ghirlanda, G},
  journal={MNRAS},
  volume={414},
  number={3},
  pages={2674--2689},
  year={2011},
  publisher={Blackwell Publishing Ltd Oxford, UK}
}

@article{ghisellini2014power,
  title={The power of relativistic jets is larger than the luminosity of their accretion disks},
  author={Ghisellini, Gabriele and Tavecchio, Fabrizio and Maraschi, L and Celotti, A and Sbarrato, T},
  journal={Nature},
  volume={515},
  number={7527},
  pages={376--378},
  year={2014},
  publisher={Nature Publishing Group UK London}
}

@article{hovatta2009doppler,
  title={Doppler factors, Lorentz factors and viewing angles for quasars, BL Lacertae objects and radio galaxies},
  author={Hovatta, Talvikki and Valtaoja, Esko and Tornikoski, Merja and L{\"a}hteenm{\"a}ki, Anne},
  journal={A\&A},
  volume={494},
  number={2},
  pages={527--537},
  year={2009},
  publisher={EDP Sciences}
}

@article{sbarrato2014jet,
  title={The jet--disc connection in AGN},
  author={Sbarrato, Tullia and Padovani, Paolo and Ghisellini, Gabriele},
  journal={MNRAS},
  volume={445},
  number={1},
  pages={81--92},
  year={2014},
  publisher={The Royal Astronomical Society}
}

@article{kutkin2019opacity,
  title={Opacity, variability, and kinematics of AGN jets},
  author={Kutkin, AM and Pashchenko, IN and Sokolovsky, KV and Kovalev, Yuri Y and Aller, MF and Aller, HD},
  journal={MNRAS},
  volume={486},
  number={1},
  pages={430--439},
  year={2019},
  publisher={Oxford University Press}
}

@article{lindfors2006synchrotron,
  title={Synchrotron flaring in the jet of 3C 279},
  author={Lindfors, Elina J and T{\"u}rler, M and Valtaoja, E and Aller, H and Aller, M and Mazin, D and Raiteri, CM and Stevens, JA and Tornikoski, M and Tosti, G and others},
  journal={A\&A},
  volume={456},
  number={3},
  pages={895--903},
  year={2006},
  publisher={EDP Sciences}
}

@article{liu2006jet,
  title={The jet power, radio loudness, and black hole mass in radio-loud active galactic nuclei},
  author={Liu, Yi and Jiang, Dong Rong and Gu, Min Feng},
  journal={ApJ},
  volume={637},
  number={2},
  pages={669},
  year={2006},
  publisher={IOP Publishing}
}

@article{liodakis2021identifying,
  title={Identifying changing jets through their radio variability},
  author={Liodakis, I and Hovatta, T and Aller, MF and Aller, HD and Gurwell, MA and L{\"a}hteenm{\"a}ki, A and Tornikoski, M},
  journal={A\&A},
  volume={654},
  pages={A169},
  year={2021},
  publisher={EDP Sciences}
}

@article{marscher1985models,
  title={Models for high-frequency radio outbursts in extragalactic sources, with application to the early 1983 millimeter-to-infrared flare of 3C 273},
  author={Marscher, Alan P and Gear, Walter Kieran},
  journal={ApJ},
  volume={298},
  pages={114--127},
  year={1985}
}

@article{okino2022collimation,
  title={Collimation of the Relativistic Jet in the Quasar 3C 273},
  author={Okino, Hiroki and Akiyama, Kazunori and Asada, Keiichi and G{\'o}mez, Jos{\'e} L and Hada, Kazuhiro and Honma, Mareki and Krichbaum, Thomas P and Kino, Motoki and Nagai, Hiroshi and Bach, Uwe and others},
  journal={ApJ},
  volume={940},
  number={1},
  pages={65},
  year={2022},
  publisher={IOP Publishing}
}

@article{paolillo2023universal,
  title={The universal shape of the X-ray variability power spectrum of AGN up to z~ 3},
  author={Paolillo, Maurizio and Papadakis, IE and Brandt, William Nielsen and Bauer, Franz E and Lanzuisi, Giorgio and Allevato, V and Shemmer, Ohad and Zheng, XC and De Cicco, D and Gilli, Roberto and others},
  journal={A\&A},
  volume={673},
  pages={A68},
  year={2023},
  publisher={EDP Sciences}
}

@article{shen2011catalog,
  title={A catalog of quasar properties from sloan digital sky survey data release 7},
  author={Shen, Yue and Richards, Gordon T and Strauss, Michael A and Hall, Patrick B and Schneider, Donald P and Snedden, Stephanie and Bizyaev, Dmitry and Brewington, Howard and Malanushenko, Viktor and Malanushenko, Elena and others},
  journal={ApJS},
  volume={194},
  number={2},
  pages={45},
  year={2011},
  publisher={IOP Publishing}
}

@article{turler1999modelling,
  title={Modelling the submillimetre-to-radio flaring behaviour of 3{C} 273},
  author={T{\"u}rler, M and Courvoisier, TJ-L and Paltani, S},
  journal={A\&A},
  volume={349},
  pages={45--54},
  year={1999}
}

@article{uttley2002measuring,
  title={Measuring the broad-band power spectra of active galactic nuclei with RXTE},
  author={Uttley, Ph and McHardy, IM and Papadakis, IE},
  journal={MNRAS},
  volume={332},
  number={1},
  pages={231--250},
  year={2002},
  publisher={Blackwell Science Ltd Oxford, UK}
}

@article{van1989quasi,
  title={Quasi-periodic oscillations and noise in low-mass X-ray binaries},
  author={Van der Klis, M},
  journal={ARA\&A},
  volume={27},
  pages={517--553},
  year={1989}
}

@article{terasranta1994brightness,
  title={Brightness temperatures and viewing angles for extragalactic radio sources: a test of unification schemes for active galactic nuclei},
  author={Terasranta, H and Valtaoja, E},
  journal={A\&A},
  volume={283},
  pages={51--58},
  year={1994}
}

@article{tseng2016structural,
  title={Structural transition in the NGC 6251 Jet: an interplay with the supermassive black hole and its host galaxy},
  author={Tseng, Chih-Yin and Asada, Keiichi and Nakamura, Masanori and Pu, Hung-Yi and Algaba, Juan-Carlos and Lo, Wen-Ping},
  journal={ApJ},
  volume={833},
  number={2},
  pages={288},
  year={2016},
  publisher={IOP Publishing}
}

@article{potter2015new,
  title={New constraints on the structure and dynamics of black hole jets},
  author={Potter, William J and Cotter, Garret},
  journal={MNRAS},
  volume={453},
  number={4},
  pages={4070--4088},
  year={2015},
  publisher={Oxford University Press}
}

@article{press1978flicker,
  title={Flicker noises in astronomy and elsewhere},
  author={Press, William H},
  journal={Comments Mod. Phys. Part C-Comments Astrophys.},
  volume={7},
  pages={103--119},
  year={1978}
}

@article{papadakis2004scaling,
  title={The scaling of the X-ray variability with black hole mass in active galactic nuclei},
  author={Papadakis, IE},
  journal={MNRAS},
  volume={348},
  number={1},
  pages={207--213},
  year={2004},
  publisher={Blackwell Science Ltd Oxford, UK}
}

@article{park2017long,
  title={The long-term centimeter variability of active galactic nuclei: A new relation between variability timescale and accretion rate},
  author={Park, Jongho and Trippe, Sascha},
  journal={ApJ},
  volume={834},
  number={2},
  pages={157},
  year={2017},
  publisher={IOP Publishing}
}

@article{pichel2023statistical,
  title={Statistical redshift of the very-high-energy blazar S5 0716+ 714},
  author={Pichel, A and Donzelli, C and Muriel, Hernan and Rovero, Adrian Carlos and Gonz{\'a}lez, D Rosa and Vega, O and Aretxaga, I and Gonz{\'a}lez, J Becerra and Terlevich, E and Terlevich, R and others},
  journal={A\&A},
  volume={680},
  pages={A52},
  year={2023},
  publisher={EDP Sciences}
}

@article{ramakrishnan2015locating,
  title={Locating the $\gamma$-ray emission site in Fermi/LAT blazars from correlation analysis between 37 GHz radio and $\gamma$-ray light curves},
  author={Ramakrishnan, V and Hovatta, T and Nieppola, E and Tornikoski, M and L{\"a}hteenm{\"a}ki, A and Valtaoja, E},
  journal={MNRAS},
  volume={452},
  number={2},
  pages={1280--1294},
  year={2015},
  publisher={The Royal Astronomical Society}
}

@article{readhead1994equipartition,
  title={Equipartition brightness temperature and the inverse Compton catastrophe},
  author={Readhead, Anthony},
  journal={ApJ},
  volume={426},
  pages={51--59},
  year={1994}
}

@article{shakura1973black,
  title={Black holes in binary systems. Observational appearance.},
  author={Shakura, Nicolai Ivanovich and Sunyaev, Rashid Alievich},
  journal={A\&A},
  volume={24},
  pages={337--355},
  year={1973}
}

@article{savolainen2002connections,
  title={Connections between millimetre continuum variations and VLBI structure in 27 AGN},
  author={Savolainen, T and Wiik, K and Valtaoja, E and Jorstad, SG and Marscher, AP},
  journal={A\&A},
  volume={394},
  number={3},
  pages={851--861},
  year={2002},
  publisher={EDP Sciences}
}

@article{shaw2012spectroscopy,
  title={Spectroscopy of Broad-line Blazars from 1LAC},
  author={Shaw, Michael S and Romani, Roger W and Cotter, Garret and Healey, Stephen E and Michelson, Peter F and Readhead, Anthony CS and Richards, Joseph L and Max-Moerbeck, Walter and King, Oliver G and Potter, William J},
  journal={ApJ},
  volume={748},
  number={1},
  pages={49},
  year={2012},
  publisher={IOP Publishing}
}

@article{stickel1993complete,
  title={The complete sample of 1 Jy BL Lac objects. II-Observational data},
  author={Stickel, M and Fried, JW and K{\"u}hr, H},
  journal={A\&AS},
  volume={98},
  pages={393--442},
  year={1993}
}

@article{valtaoja1992five,
  title={Five years monitoring of extragalactic radio sources-part three-generalized shock models and the dependence of variability on frequency},
  author={Valtaoja, E and Terasranta, H and Urpo, S and Nesterov, NS and Lainela, M and Valtonen, M},
  journal={A\&A},
  volume={254},
  pages={71},
  year={1992}
}

@article{hovatta2007statistical,
  title={Statistical analyses of long-term variability of AGN at high radio frequencies},
  author={Hovatta, T and Tornikoski, M and Lainela, M and Lehto, HJ and Valtaoja, E and Torniainen, I and Aller, MF and Aller, HD},
  journal={A\&A},
  volume={469},
  number={3},
  pages={899--912},
  year={2007},
  publisher={EDP Sciences}
}

@article{terasranta1998fifteen,
  title={Fifteen years monitoring of extragalactic radio sources at 22, 37 and 87 GHz},
  author={Ter{\"a}sranta, Harri and Tornikoski, M and Mujunen, A and Karlamaa, K and Valtonen, T and Henelius, N and Urpo, S and Lainela, M and Pursimo, T and Nilsson, K and others},
  journal={A\&AS},
  volume={132},
  number={3},
  pages={305--331},
  year={1998},
  publisher={EDP Sciences}
}

@article{uhlenbeck1930theory,
  title={On the theory of the Brownian motion},
  author={Uhlenbeck, George E and Ornstein, Leonard S},
  journal={Phys. Rev.},
  volume={36},
  number={5},
  pages={823},
  year={1930},
  publisher={APS}
}

@article{torrealba2012optical,
  title={Optical spectroscopic atlas of the MOJAVE/2cm AGN sample},
  author={Torrealba, Janet and Chavushyan, Vahram and Cruz-Gonz{\'a}lez, Irene and Arshakian, Tigran G and Bertone, Emanuele and Rosa-Gonzalez, Daniel},
  journal={Rev. Mex. Astron. Astrofis.},
  volume={48},
  number={1},
  pages={09--40},
  year={2012},
  publisher={Instituto de Astronom{\'\i}a}
}

@article{weaver2022kinematics,
  title={Kinematics of Parsec-scale Jets of Gamma-Ray Blazars at 43 GHz during 10 yr of the VLBA-BU-BLAZAR Program},
  author={Weaver, Zachary R and Jorstad, Svetlana G and Marscher, Alan P and Morozova, Daria A and Troitsky, Ivan S and Agudo, Iv{\'a}n and G{\'o}mez, Jos{\'e} L and L{\"a}hteenm{\"a}ki, Anne and Tammi, Joni and Tornikoski, Merja},
  journal={ApJS},
  volume={260},
  number={1},
  pages={12},
  year={2022},
  publisher={IOP Publishing}
}

@article{vlahakis2003relativistic,
  title={Relativistic magnetohydrodynamics with application to gamma-ray burst outflows. I. Theory and semianalytic trans-Alfv{\'e}nic solutions},
  author={Vlahakis, Nektarios and K{\"o}nigl, Arieh},
  journal={ApJ},
  volume={596},
  number={2},
  pages={1080},
  year={2003},
  publisher={IOP Publishing}
}

@article{wilkinson2009accretion,
  title={Accretion disc variability in the hard state of black hole X-ray binaries},
  author={Wilkinson, Tony and Uttley, Philip},
  journal={MNRAS},
  volume={397},
  number={2},
  pages={666--676},
  year={2009},
  publisher={Blackwell Publishing Ltd Oxford, UK}
}

@article{zamaninasab2014dynamically,
  title={Dynamically important magnetic fields near accreting supermassive black holes},
  author={Zamaninasab, Mohammad and Clausen-Brown, E and Savolainen, T and Tchekhovskoy, A},
  journal={Nature},
  volume={510},
  number={7503},
  pages={126--128},
  year={2014},
  publisher={Nature Publishing Group UK London}
}

\begin{appendix}
\onecolumn
\section{Tables}


\renewcommand{\arraystretch}{1.3}
\begin{longtable}[h!]{lllllllllll} 
\caption{\label{SourcesDopplersLorentz} Variability Doppler factors, jet bulk Lorentz factors, and central-engine parameters.}\\
\hline\hline
Source	&	Alias & Type	& $z$ &\(\delta_{\rm var}\)&	\(\beta_{\rm app}\) (c) & \(\Gamma\) & log \(L_{\rm acc}\) (erg/s) & log\(M_{\rm BH}\)(\(M\odot\))  & \(\dot m\) & ref \\
\hline

\endfirsthead
\caption{Continued.}\\
\hline \hline
Source	&	Alias & Type	& $z$ &\(\delta_{\rm var}\)&	\(\beta_{\rm app}\) (c) & \(\Gamma\) & log \(L_{\rm acc}\) (erg/s) & log\(M_{\rm BH}\)(\(M\odot\))  & \(\dot m\) & ref \\
\hline
\endhead
\hline      
\hline
0007+106	&	PG0007+106	&	GAL	&	0.089	&	$2.9_{-0.6}^{+0.6}$	&	1.7	&	2.1	&	$45.6_{-0.8}^{+0.7}$	&	8.2	&	$0.43_{-0.35}^{+1.94}$	&	Tor12	\\
0059+581	&	0059+581	&	FSRQ	&	0.644	&	$20.0_{-4.5}^{+5.2}$	&	8.6	&	11.9	&	$46.3_{-0.7}^{+0.7}$	&	8.4	&	$1.60_{-1.28}^{+6.23}$	&	Sha12	\\
0106+013	&	0106+013	&	FSRQ	&	2.110	&	$18.4_{-5.4}^{+5.9}$	&	25.9	&	27.5	&		&		&		&		\\
0109+22	&	S20109+22	&	BLO	&	0.49	&	$8.7_{-1.7}^{+2.6}$	&	0.3	&	4.4	&		&		&		&		\\
0133+476	&	0133+476	&	FSRQ	&	0.859	&	$19.7_{-5.2}^{+4.4}$	&	17.0	&	17.2	&	$46.1_{-0.8}^{+0.8}$	&	8.7	&	$0.45_{-0.37}^{+2.08}$	&	Tor12	\\
0234+285	&	0234+285	&	FSRQ	&	1.206	&	$14.8_{-5.5}^{+7.4}$	&	25.1	&	28.7	&	$46.8_{-0.7}^{+0.7}$	&	8.9	&	$1.43_{-1.14}^{+5.72}$	&	Sha12	\\
0235+164	&	0235+164	&	BLO	&	0.94	&	$26.2_{-8.8}^{+9.9}$	&	26.9	&	27.0	&	$45.9_{-0.7}^{+0.7}$	&	7.9	&	$2.11_{-1.68}^{+8.22}$	&	Coh87	\\
0316+413	&	3C84	&	GAL	&	0.018	&		&	0.43	&		&		&		&		&\\
0333+321	&	0333+321	&	FSRQ	&	1.259	&	$11.8_{-4.5}^{+4.5}$	&	13.4	&	13.6	&	$47.9_{-0.7}^{+0.7}$	&	9.4	&	$5.55_{-4.46}^{+23.56}$	&	Tor12	\\
0336-019	&	CTA026	&	FSRQ	&	0.848	&	$15.0_{-3.6}^{+4.4}$	&	25.0	&	28.4	&	$46.5_{-0.7}^{+0.7}$	&	8.7	&	$1.30_{-1.04}^{+5.22}$	&	Tor12	\\
0355+508	&	0355+508	&	FSRQ	&	1.52	&	$17.8_{-7.4}^{+10.7}$	&	8.8	&	11.1	&		&		&		&		\\
0415+379	&	0415+379	&	GAL	&	0.049	&	$3.2_{-0.7}^{+0.5}$	&	7.9	&	11.4	&	$45.8_{-0.8}^{+0.8}$	&	8.3	&	$0.65_{-0.54}^{+3.01}$	&	Tor12	\\
0420-014	&	0420-014	&	FSRQ	&	0.914	&	$13.3_{-3.5}^{+7.3}$	&	5.6	&	7.8	&	$46.4_{-0.7}^{+0.7}$	&	8.6	&	$1.21_{-0.97}^{+4.87}$	&	Tor12	\\
0422+0036	&	PKS0422+0036	&	BLO	&	0.268	&	$2.8_{-0.6}^{+0.8}$	&	0.8	&	1.7	&		&		&		&		\\
0430+052	&	3C120	&	GAL	&	0.033	&	$3.3_{-0.6}^{+1.0}$	&	6.6	&	8.3	&	$45.2_{-0.7}^{+0.7}$	&	7.5	&	$0.84_{-0.69}^{+3.77}$	&	Tor12	\\
0528+134	&	0528+134	&	FSRQ	&	2.07	&	$23.0_{-10.0}^{+13.5}$	&	18.6	&	19.0	&		&		&		&		\\
0552+398	&	0552+398	&	FSRQ	&	2.363	&	$18.7_{-6.0}^{+11.2}$	&	1.3	&	9.4	&		&		&		&		\\
0642+449	&	0642+449	&	FSRQ	&	3.396	&	$25.7_{-8.1}^{+10.4}$	&	8.6	&	14.3	&		&		&		&		\\
0716+714	&	0716+714	&	BLO	&	0.230	&	$23.9_{-4.0}^{+5.5}$	&	34.5	&	36.9	&		&		&		&		\\
0735+17	&	PKS0735+17	&	BLO	&	0.424	&	$5.7_{-2.5}^{+3.9}$	&	6.9	&	7.1	&		&		&		&		\\
0736+017	&	0736+017	&	FSRQ	&	0.189	&	$5.5_{-0.8}^{+1.0}$	&	12.5	&	16.9	&	$45.6_{-0.8}^{+0.8}$	&	7.9	&	$1.01_{-0.83}^{+4.67}$	&	Tor12	\\
0804+499	&	0804+499	&	FSRQ	&	1.436	&	$19.6_{-3.0}^{+3.6}$	&	1.0	&	9.8	&	$46.5_{-0.7}^{+0.7}$	&	9.0	&	$0.53_{-0.42}^{+2.06}$	&	Tor12	\\
0827+243	&	OJ248	&	FSRQ	&	0.941	&	$16.5_{-4.5}^{+5.1}$	&	20.3	&	20.8	&	$46.4_{-0.7}^{+0.7}$	&	8.7	&	$1.01_{-0.81}^{+4.05}$	&	Tor12	\\
0836+710	&	0836+710	&	FSRQ	&	2.198	&	$78.6_{-11.8}^{+12.3}$	&	21.7	&	42.3	&		&		&		&		\\
0851+202	&	OJ287	&	BLO	&	0.306	&	$16.9_{-2.5}^{+7.6}$	&	15.8	&	15.9	&	$45.7_{-0.7}^{+0.7}$	&		&		&	Sti93	\\
0923+392	&	4C39.25	&	FSRQ	&	0.696	&	$1.8_{-0.7}^{+1.2}$	&	2.7	&	3.3	&	$46.2_{-0.8}^{+0.8}$	&	8.8	&	$0.45_{-0.37}^{+2.14}$	&	Tor12	\\
0953+254	&	0953+254	&	FSRQ	&	0.708	&	$5.0_{-1.1}^{+1.2}$	&	10.4	&	13.3	&	$46.4_{-0.7}^{+0.7}$	&	8.5	&	$1.39_{-1.11}^{+5.59}$	&	Tor12	\\
1055+018	&	1055+018	&	FSRQ	&	0.892	&	$17.5_{-4.8}^{+4.2}$	&	6.8	&	10.1	&	$46.4_{-0.7}^{+0.7}$	&	8.8	&	$0.75_{-0.60}^{+3.00}$	&	Tor12	\\
1156+295	&	4C29.45	&	FSRQ	&	0.725	&	$18.1_{-5.8}^{+7.6}$	&	25.4	&	26.8	&	$46.3_{-0.7}^{+0.7}$	&	8.6	&	$1.08_{-0.86}^{+4.34}$	&	Tor12	\\
1219+285	&	ON231	&	BLO	&	0.102	&		&	8.6	&	&		&		&		&		\\
1222+216	&	PKS1222+216	&	FSRQ	&	0.433	&	$5.1_{-1.9}^{+3.5}$	&	22.7	&	52.6	&	$46.2_{-0.8}^{+0.8}$	&	9.1	&	$0.25_{-0.20}^{+1.15}$	&	Tor12	\\
1226+023	&	3C273	&	FSRQ	&	0.158	&	$5.1_{-1.3}^{+1.6}$	&	15.5	&	26.5	&	$46.6_{-0.8}^{+0.8}$	&	9.1	&	$0.59_{-0.49}^{+2.83}$	&	Tor12	\\
1253-055	&	3C279	&	FSRQ	&	0.536	&	$10.9_{-1.2}^{+2.7}$	&	21.2	&	26.1	&		&		&		&		\\
1308+326	&	1308+326	&	FSRQ	&	0.995	&	$11.3_{-3.1}^{+4.6}$	&	28.1	&	40.7	&	$46.2_{-0.7}^{+0.7}$	&	8.6	&	$0.90_{-0.72}^{+3.61}$	&	Tor12	\\
1413+135	&	PKS1413+135	&	BLO	&	0.247	&	$5.9_{-1.0}^{+1.0}$	&	1.8	&	3.3	&		&		&		&		\\
1418+546	&	OQ530	&	BLO	&	0.152	&	$2.6_{-0.5}^{+0.6}$	&	1.0	&	1.7	&	$45.5_{-0.7}^{+0.7}$	&		&		&	Sti93	\\
1502+106	&	PKS1502+106	&	FSRQ	&	1.834	&	$17.4_{-6.4}^{+8.5}$	&	17.5	&	17.6	&		&		&		&		\\
1510-089	&	PKS1510-089	&	FSRQ	&	0.36	&	$18.4_{-3.2}^{+5.1}$	&	29.2	&	32.4	&	$46.1_{-0.8}^{+0.8}$	&	8.4	&	$0.92_{-0.76}^{+4.26}$	&	Tor12	\\
1606+106	&	1606+106	&	FSRQ	&	1.235	&	$14.2_{-2.8}^{+4.3}$	&	18.3	&	19.0	&	$46.5_{-0.7}^{+0.7}$	&	8.7	&	$1.24_{-0.99}^{+4.98}$	&	Tor12	\\
1611+343	&	DA406	&	FSRQ	&	1.4	&	$11.6_{-4.9}^{+7.5}$	&	31.7	&	49.0	&	$46.7_{-0.7}^{+0.7}$	&	8.9	&	$1.11_{-0.89}^{+4.44}$	&	Tor12	\\
1633+382	&	4C38.41	&	FSRQ	&	1.815	&	$30.6_{-8.3}^{+10.5}$	&	31.3	&	31.3	&	$46.6_{-0.7}^{+0.7}$	&	8.9	&	$1.06_{-0.85}^{+4.24}$	&	Tor12	\\
1637+574	&	1637+574	&	FSRQ	&	0.751	&	$12.9_{-2.6}^{+3.3}$	&	11.5	&	11.6	&	$46.4_{-0.8}^{+0.8}$	&	8.8	&	$0.84_{-0.69}^{+3.99}$	&	Tor12	\\
1641+399	&	3C345	&	FSRQ	&	0.594	&	$7.6_{-2.5}^{+7.0}$	&	20.1	&	30.3	&	$45.9_{-0.8}^{+0.8}$	&	8.4	&	$0.57_{-0.47}^{+2.62}$	&	Tor12	\\
1730-130	&	1730-130	&	FSRQ	&	0.902	&	$17.7_{-4.3}^{+6.2}$	&	28.1	&	31.1	&		&		&		&		\\
1741-038	&	1741-038	&	FSRQ	&	1.054	&	$16.1_{-4.5}^{+5.1}$	&		&	&		&		&		&		\\
1749+096	&	PKS1749+096	&	BLO	&	0.322	&	$13.2_{-1.8}^{+2.1}$	&	7.1	&	8.6	&	$45.6_{-0.7}^{+0.8}$	&		&		&	Sti93	\\
2005+403	&	2005+403	&	FSRQ	&	1.736	&	$16.7_{-13.1}^{+7.6}$	&	9.9	&	11.3	&		&		&		&		\\
2144+092	&	2144+092	&	FSRQ	&	1.113	&	$6.7_{-1.6}^{+1.9}$	&	0.9	&	3.5	&		&		&		&		\\
2145+067	&	2145+067	&	FSRQ	&	0.999	&	$11.8_{-3.0}^{+7.4}$	&	3.2	&	6.4	&	$47.1_{-0.7}^{+0.7}$	&	9.2	&	$1.53_{-1.23}^{+6.31}$	&	Tor12	\\
2200+420	&	BL Lac	&	BLO	&	0.069	&	$4.7_{-0.8}^{+1.8}$	&	10.5	&	14.3	&	$45.2_{-0.7}^{+0.7}$	&		&		&	Sti93	\\
2201+315	&	2201+315	&	FSRQ	&	0.295	&	$7.8_{-2.0}^{+2.5}$	&	9.0	&	9.2	&	$45.5_{-0.7}^{+0.8}$	&	8.2	&	$0.38_{-0.31}^{+1.77}$	&	Tor12	\\
2223-052	&	3C446	&	FSRQ	&	1.404	&	$15.5_{-5.9}^{+9.7}$	&	18.1	&	18.3	&		&		&		&		\\
2230+114	&	2230+114	&	FSRQ	&	1.038	&	$22.5_{-5.6}^{+8.0}$	&	20.5	&	20.6	&	$46.6_{-0.7}^{+0.7}$	&	8.7	&	$1.53_{-1.22}^{+6.13}$	&	Tor12	\\
2251+158	&	3C454.3	&	FSRQ	&	0.859	&	$22.5_{-6.5}^{+10.5}$	&	17.5	&	18.1	&	$46.8_{-0.7}^{+0.7}$	&	8.9	&	$1.33_{-1.06}^{+5.34}$	&	Tor12	\\
\hline
\hline \label{table:sources} 

\end{longtable}
\tablefoot{The references for the line luminosities (and FWHMs when available) are Tor12: \cite{torrealba2012optical}, Sha12: \cite{shaw2012spectroscopy}, Sti93: \cite{stickel1993complete}, Coh87: \cite{cohen1987nature}. The apparent speeds in units of $\rm \mu as/y$ and the redshifts, apart from the redshifts for 0716+714 (\citealt{pichel2023statistical}) and S20109+22 (\citealt{koljonen2024galaxy}), were obtained from the MOJAVE database, maintained by the MOJAVE team (\citealt{lister2018mojave}). The median Doppler factor \(\delta_{\rm var}\) is given with the 68 \% confidence interval. For \(L_{\rm acc}\) and \(\dot m\) the limits were obtained from Eq. \ref{Hbeta}-\ref{OIII}.}

\renewcommand{\arraystretch}{1.3}
\begin{longtable}[h!]{lllllllll}
\caption{\label{timescales} Obtained timescales.}\\
\hline\hline
Source	&		\(T_{\rm PSD}\)	& \(T_{\rm dur,obs}\) &\(T_{\rm sep,obs}\) & \(\tau_{\rm rise,obs}\)\\
\hline
\endfirsthead
\caption{Continued.}\\
\hline \hline
Source	&		\(T_{\rm PSD}\)	& \(T_{\rm dur,obs}\) &\(T_{\rm sep,obs}\) & \(\tau_{\rm rise,obs}\)\\
\hline
\endhead
\hline      
\hline
\textbf{0007+106}	&	$1000_{-200}^{+1000}$	&	$828\pm115$	&	$487\pm124$	&	$83_{-22}^{+54}$		\\
0059+581	&	$6000_{-4500}^{>1000}$	&	$1198\pm145$	&	$478\pm60$	&	$95_{-26}^{+57}$		\\
0106+013	&	$7000_{-4500}^{>}$	&	$3436\pm564$	&	$1233\pm209$	&	$368_{-128}^{+249}$		\\
0109+22	&	$6500_{-5000}^{>500}$	&	$832\pm214$	&	$507\pm113$	&	$88_{-33}^{+42}$		\\
0133+476	&	$6000_{-4500}^{>1000}$	&	$2103\pm310$	&	$820\pm105$	&	$54_{-11}^{+25}$		\\
0234+285	&	$5000_{-4200}^{>2000}$	&	$2246\pm411$	&	$942\pm221$	&	$220_{-149}^{+233}$		\\
\textbf{0235+164}	&	$1000_{-200}^{+500}$	&	$874\pm142$	&	$699\pm65$	&	$144_{-23}^{+11}$		\\
0316+413	&	$7000_{-1000}^{>}$	&	$13683\pm72$	&	$12093\pm16$	&	$1537_{-25}^{+21}$		\\
0333+321	&	$3000_{-2200}^{>4000}$	&	$2730\pm358$	&	$775\pm156$	&	$409_{-112}^{+95}$		\\
0336-019	&	$5500_{-4700}^{>1500}$	&	$1397\pm624$	&	$763\pm101$	&	$72_{-30}^{+45}$		\\
0355+508	&	$7000_{-4500}^{>}$	&	$4463\pm1182$	&	$2375\pm665$	&	$649_{-575}^{+292}$		\\
\textbf{0415+379}	&	$1500_{-500}^{+2500}$	&	$1261\pm295$	&	$696\pm75$	&	$60_{-20}^{+8}$		\\
0420-014	&	$7000_{-3500}^{>}$	&	$3935\pm332$	&	$1201\pm232$	&	$350_{-30}^{+409}$		\\
0422+0036	&	$1000_{-600}^{>6000}$	&	$1299\pm326$	&	$524\pm178$	&	$146_{-56}^{+113}$		\\
\textbf{0430+052}	&	$2500_{-1000}^{+3000}$	&	$512\pm69$	&	$1051\pm139$	&	$41_{-7}^{+11}$		\\
0528+134	&	$7000_{-5000}^{>}$	&	$2158\pm232$	&	$813\pm124$	&	$323_{-75}^{+66}$		\\
0552+398	&		&	$2358\pm827$	&	$805\pm337$	&	$879_{-618}^{+1439}$		\\
0642+449	&	$7000_{-4500}^{>}$	&	$2377\pm1547$	&	$1943\pm680$	&	$101_{-55}^{+167}$		\\
\textbf{0716+714}	&	$500_{-100}^{+500}$	&	$339\pm29$	&	$342\pm18$	&	$39_{-5}^{+110}$		\\
0735+17	&	$7000_{-4000}^{>}$	&	$3186\pm712$	&	$1239\pm462$	&	$276_{-85}^{+163}$		\\
\textbf{0736+017}	&	$1000_{-500}^{+1000}$	&	$575\pm75$	&	$523\pm101$	&	$61_{-20}^{+46}$		\\
0804+499	&	$6500_{-5500}^{>500}$	&	$520\pm134$	&	$598\pm161$	&	$35_{-12}^{+31}$		\\
0827+243	&	$2500_{-2000}^{>4500}$	&	$959\pm145$	&	$528\pm85$	&	$87_{-25}^{+58}$		\\
0836+710	&	$7000_{-5000}^{>}$	&	$635\pm81$	&	$426\pm42$	&	$7_{-2}^{+5}$		\\
0851+202	&	$3000_{-2200}^{>4000}$	&	$814\pm40$	&	$864\pm49$	&	$72_{-12}^{+6}$		\\
0923+392	&		&	$12009\pm3696$	&	$4054\pm1085$	&	$2232_{-312}^{+196}$		\\
0953+254	&	$3000_{-2600}^{>4000}$	&	$1804\pm678$	&	$899\pm217$	&	$176_{-48}^{+211}$		\\
1055+018	&	$7000_{-5500}^{>}$	&	$1895\pm993$	&	$1104\pm159$	&	$140_{-65}^{+129}$		\\
\textbf{1156+295}	&	$1500_{-500}^{+1500}$	&	$1698\pm187$	&	$1082\pm131$	&	$127_{-27}^{+10}$		\\
1219+285	&	$6500_{-5000}^{>500}$	&	$2042\pm964$	&	$1371\pm614$	&	$431_{-361}^{+654}$		\\
1222+216	&	$7000_{-4500}^{>}$	&	$3148\pm625$	&	$1364\pm222$	&	$503_{-202}^{+127}$		\\
1226+023	&	$7000_{-5000}^{>}$	&	$2522\pm112$	&	$1454\pm115$	&	$240_{-45}^{+15}$		\\
1253-055	&	$4500_{-2500}^{>2500}$	&	$2566\pm325$	&	$1107\pm109$	&	$144_{-40}^{+12}$		\\
1308+326	&	$6500_{-3000}^{>500}$	&	$3329\pm1065$	&	$971\pm272$	&	$374_{-177}^{+87}$		\\
1413+135	&	$7000_{-5500}^{>}$	&	$808\pm155$	&	$421\pm58$	&	$68_{-28}^{+62}$		\\
1418+546	&	$6500_{-0}^{>500}$	&	$1164\pm375$	&	$698\pm193$	&	$159_{-50}^{+93}$		\\
1502+106	&	$6000_{-4000}^{>1000}$	&	$2179\pm587$	&	$951\pm169$	&	$362_{-118}^{+151}$		\\
1510-089	&	$6000_{-5600}^{>1000}$	&	$784\pm119$	&	$625\pm84$	&	$46_{-11}^{+8}$		\\
1606+106	&	$7000_{-4500}^{>}$	&	$944\pm237$	&	$395\pm82$	&	$74_{-24}^{+53}$		\\
1611+343	&	$7000_{-4500}^{>}$	&	$3033\pm739$	&	$1208\pm282$	&	$249_{-61}^{+155}$		\\
1633+382	&	$6000_{-3500}^{>1000}$	&	$2541\pm879$	&	$1315\pm288$	&	$154_{-81}^{+67}$		\\
1637+574	&	$6000_{-4500}^{>1000}$	&	$2269\pm968$	&	$729\pm335$	&	$57_{-19}^{+120}$		\\
1641+399	&	$7000_{-500}^{>}$	&	$4240\pm588$	&	$1873\pm407$	&	$374_{-50}^{+242}$		\\
1730-130	&	$7000_{-5000}^{>}$	&	$1252\pm402$	&	$731\pm167$	&	$556_{-280}^{+144}$		\\
1741-038	&	$6000_{-4000}^{>1000}$	&	$2270\pm486$	&	$497\pm133$	&	$210_{-67}^{+149}$		\\
\textbf{1749+096}	&	$1000_{-200}^{+500}$	&	$1126\pm116$	&	$687\pm54$	&	$183_{-105}^{+87}$		\\
2005+403	&		&	$5672\pm1212$	&	$2865\pm984$	&	$259_{-229}^{+561}$		\\
\textbf{2144+092}	&	$800_{-400}^{+1200}$	&	$1098\pm666$	&	$535\pm196$	&	$96_{-33}^{+62}$		\\
2145+067	&		&	$6655\pm974$	&	$1948\pm332$	&	$195_{-69}^{+81}$		\\
2200+420	&	$7000_{-5000}^{>}$	&	$1589\pm153$	&	$1187\pm112$	&	$94_{-25}^{+28}$		\\
2201+315	&	$5000_{-3500}^{>2000}$	&	$2509\pm762$	&	$1876\pm637$	&	$95_{-46}^{+27}$		\\
2223-052	&	$7000_{-2500}^{>}$	&	$2870\pm448$	&	$736\pm203$	&	$934_{-323}^{+439}$		\\
\textbf{2230+114}	&	$1000_{-200}^{+500}$	&	$1069\pm153$	&	$1045\pm158$	&	$152_{-17}^{+33}$		\\
\textbf{2251+158}	&	$3000_{-1000}^{>4000}$	&	$1530\pm141$	&	$1621\pm772$	&	$140_{-5}^{+4}$		\\
\hline
\hline \label{table:timescales} 
\end{longtable}
\tablefoot{$T_{\rm PSD}$ is the PSD timescale obtained in Paper I, $T_{\rm dur,obs}$ is the mean flare duration, $T_{\rm sep,obs}$ is the mean flare separation and $\tau_{\rm rise,obs}$ is the median rise time. The 90 \% confidence limits for $T_{PSD}$ of constrained sources (in bold) include the limits given by the best-fit PSD slope. For the other sources, the confidence limits give any $T_{\rm PSD}$ in the parameter space of possible slopes. The symbol $>$ in $T_{\rm PSD}$ indicates that the upper confidence limit is at the maximum timescale analysed in Paper I and thus may be longer. Both $T_{\rm dur,obs}$ and  $T_{\rm sep,obs}$ are given with standard deviations, and $\tau_{\rm rise,obs}$ includes the 68 \% confidence interval. The timescales are all given in days. }

\end{appendix}
\end{document}